\documentclass[]{JHEP}
\usepackage{epsfig}

\def\bea{\begin{eqnarray}}
\def\eea{\end{eqnarray}}
\def\be{\begin{equation}}
\def\ee{\end{equation}}
\def\nn{\nonumber}

\def\Z{{\bf Z}}
\def\m{\raisebox{-0.6pt}{\mbox{-}}\hspace{-0.5pt}}
\def\mm{\raisebox{-0.9pt}{\mbox{-}}\hspace{-0.5pt}}
\def\dl{\scriptscriptstyle{d_l}}
\def\Tb{\hspace{1pt}\overline{\hspace{-1pt} T}\hspace{-1pt}}
\def\psib{\hspace{3pt}\overline{\hspace{-3pt} \psi}\hspace{0pt}}
\def\Xb{\hspace{2pt}\overline{\hspace{-3pt} X \hspace{-1pt}}\hspace{1pt}}

\newcommand{\eqalign}[1]{\hspace{-10pt}\begin{array}{ll}#1\end{array}
\hspace{-10pt}}

\title{Target-space anomalies and elliptic indices
in heterotic orbifolds}

\author{Claudio A. Scrucca \\
CERN, 1211 Geneva 23, Switzerland \\
{\footnotesize \tt Claudio.Scrucca@cern.ch}}
\author{Marco Serone \\
ISAS-SISSA, Via Beirut 2-4, 34013 Trieste, Italy \\
INFN, sez. di Trieste, Italy \\
{\footnotesize \tt serone@sissa.it}}

\abstract{We present a complete string theory analysis of all mixed
gauge, gravitational and target-space anomalies potentially arising 
in the simplest heterotic $\Z_N$ orbifold models, with $N$ odd and 
standard embedding. These anomalies turn out to be encoded in an 
elliptic index, which can be easily computed; they are found to 
cancel through a universal GS mechanism induced by the dilaton multiplet. 
The target-space symmetry is then shown to have a nice geometric 
interpretation in terms of torsion, and the target-space dependence of 
the four-dimensional GS couplings can be alternatively rederived from 
the implicit torsion dependence of the standard ten-dimensional GS 
couplings. The result is universal and consists essentially of a 
Bianchi identity for the NSNS $B$ field depending on all the curvatures, 
and in particular on the target-space curvature. 
}

\preprint{CERN-TH/2000-364 \\ SISSA-112/2000/EP \\ 
{\tt hep-th/0012124}}

\begin{document}


\section{Introduction}

Target-space duality symmetries (see \cite{gpr} and references
therein) in orbifold compactifications of heterotic string theory 
\cite{dhvw} have received a lot of attention in the past and have 
recently been the object of renewed interest \cite{iru2}--\cite{ss4} 
in the context of heterotic--Type I duality \cite{polwit}, which has been
conjectured to exist also between certain $D=4$ heterotic and Type IIB 
orientifold vacua \cite{abpss}--\cite{afiuv}.
At the level of the low-energy supergravity effective action, these 
symmetries correspond typically to global isometries acting on the 
manifold describing the moduli space of scalar fields. An example is 
given by $D=4$ $N=1$ heterotic orbifold models, which admit $SL(2,R)_i$ 
target-space duality symmetries acting in a peculiar way on the untwisted 
moduli $T_i$ that describe the K\"ahler structure of the internal 
space \cite{DVV}. 
Such symmetries are in general affected by one-loop anomalies\footnote{The
presence of anomalies in this kind of symmetries was first addressed in
\cite{mn} and their possible cancellation through a generalized
GS mechanism noted in \cite{HuWi}.} \cite{dkl}--\cite{bg2}, 
but since a discrete version of them (T-duality) 
is believed to be an exact symmetry of heterotic string theory, 
the latter are expected to cancel by some mechanism.

Although anomalies in target-space duality symmetries (in the following
denoted simply target-space anomalies) can arise in different channels,
most of the attention in the past has been devoted to mixed target-space/gauge
and target-space/gravitational anomalies for their relation to threshold 
corrections. It was realized that such field theory one-loop anomalies 
can be generically cancelled by the combination of a universal Green--Schwarz 
(GS) mechanism \cite{GS2}--\cite{dsw} involving the dilaton multiplet 
and the anomalous variation 
of threshold corrections to gauge couplings. The presence of a GS coupling 
has also been confirmed in \cite{agnt} by a string computation. Other cases, 
such as purely target-space anomalies, have received little attention to date,
and as far as we know, have been analysed only in \cite{bg2}, 
from a low-energy field theory point of view. 
It should be noticed that in all previous analyses,
target-space anomalies have always been computed indirectly, by exploiting 
their similarity with gauge anomalies. The difficulty of a direct computation 
resides in the fact that the connection associated to a target-space duality 
symmetry is a composite rather than an elementary field.

In this paper, we reconsider this issue through an explicit string theory 
computation of all possible target-space/gauge/gravitational anomalies 
in heterotic orbifold models. We restrict to models without threshold 
corrections, where anomalies are expected to be cancelled through a GS 
mechanism only, and focus on the simplest $\Z_N$ models with 
$N$ odd and standard embedding, i.e. the $\Z_3$ and the $\Z_7$ models. 
Anomalous amplitudes in heterotic string theory have been extensively 
studied in the past. In this work, we proceed along the lines of 
\cite{aw}--\cite{lerche} and \cite{kut}--\cite{ya} to identify the 
generalizations of the elliptic genus \cite{w2} that are relevant to the 
anomalies under consideration. By evaluating both the modular-invariant 
and holomorphic versions of the elliptic genus, we are then able to derive 
both the one-loop field theory anomaly and the GS couplings. 
We show that the one-loop anomaly nicely factorizes and is completely 
cancelled by a GS mechanism. In a low-energy description in terms of 
chiral multiplets, the latter is induced by an inhomogeneous transformation 
of the dilaton multiplet, and the cancellation of pure target-space 
anomalies requires an appropriate kinetic term for the composite 
target-space connection in the low-energy effective action, as already 
conjectured in \cite{bg2}. In a dual description in terms of linear 
multiplets, this term modifies the Bianchi identity for the NSNS 
two-form $B$ dual to the universal axion by a source term involving 
the target-space curvature.

The results we find agree with those derived in \cite{dfkz,il} 
for target-space/gauge and target-space/gravitational anomalies, and with 
the various modular weights assignments of \cite{il}. However, a disagreement 
with the results of \cite{bg2} for pure target-space anomalies is found. 
This suggests that the analogy between target-space and gauge anomalies 
used in \cite{bg2} could present subtleties related to the compositeness 
of the target-space connection, and might be incorrect when more than 
one composite connection occurs as external states. 
In this respect, we have checked that the anomaly computed 
along the lines of \cite{bg2} does not factorize and hence could not
be cancelled by a GS mechanism. 

We also clarify the geometrical structure of target-space anomalies
by showing that the four-dimensional target-space curvatures are nothing 
but the internal components of the ten-dimensional torsion-full curvature 
two-form. The additional term in the Bianchi identity for $B$ can then be 
deduced from the implicit torsion dependence of the ten-dimensional GS 
term, pointed out in \cite{hull2}. A nice geometric interpretation can 
also be made of the contribution of each single state to the anomaly in 
our string set-up. In particular, target-space anomalies resemble standard 
gravitational and gauge anomalies in the untwisted and twisted sectors 
of the orbifold, respectively. This is explained by the quite different 
dependence on the internal metric and volume in the two cases.

The structure of this paper is as follows.
In section 2, we review well-known general properties of anomalous
heterotic amplitudes. In section 3, we describe our strategy for the
string computation and hence compute the elliptic genus relevant to all 
target-space/gauge/gravitational anomalies. Some comments on the field-theory 
interpretation of our results are then given in section 4, whereas the 
geometric structure of these anomalies, including their relation with torsion, 
is given in section 5. Finally, section 6 contains some conclusions.
In addition, we report some conventions and several details
of our computation in four appendices.

\section{Anomalies in heterotic string theory}

In string theory, anomalous amplitudes happen to be total modular 
derivatives, and therefore receives contributions only from the boundaries 
of the moduli space of the relevant world-sheet surface. 
In heterotic models, for instance, anomalies are expressed as integrals
over the boundary $\partial {\cal F}$ of the fundamental domain 
${\cal F}$ of toroidal world-sheets, and their cancellation is then 
a direct consequence of modular invariance. This has been shown in detail 
in \cite{kut}--\cite{ya} for $D=10$ heterotic theories. 

The arguments of \cite{kut}--\cite{ya} can easily be generalized to 
$D=4$ vacua. In order to do so, it is convenient to regard 
the anomaly as a possible non-vanishing amplitude involving an
unphysical longitudinally polarized particle. The relevant string 
world-sheet is a torus. Moreover, since standard chiral anomalies 
only arise in the CP-odd part of the effective action, whereas others,
such as target-space anomalies, distribute in a supersymmetric way between
CP-even and CP-odd parts, it will be sufficient to restrict to correlations 
in the odd spin structure. Recall that in this completely periodic 
spin structure there is a world-sheet gravitino zero-mode inducing the 
insertion of a world-sheet supercurrent, the picture-changing operator
$T_F$. Moreover, owing to the presence of a Killing spinor, one of the 
vertex operators must be taken in the so-called $(-1)$-picture, while
all the others can be taken in the $0$-picture \cite{fms}. It is easy to 
verify that anomalous amplitudes can arise only when the vertex operator 
associated to the longitudinal particle is the one in the $(-1)$-picture. 
However, the unphysical $(-1)$-picture vertex $V^{\rm unphy.}$ is BRS-trivial, 
and can be rewritten as $Q \cdot \hat V^{\rm unphy.}$ for some appropriate 
$\hat V^{\rm unphy.}$. One can then bring $Q$ to act on the rest of the 
correlation. Since the action of $Q$ on the physical $0$-picture vertices 
$V^{\rm phy.}$ is trivial, its only effect is to convert $T_F$ into the 
right-moving world-sheet energy--momentum tensor $T_B$: $Q \cdot T_F = T_B$. 
Finally, the net effect of the insertion of $T_B$ is to produce the
derivative with respect to the torus modulus $\bar \tau$ of the remaining 
correlation\footnote{For convenience, we take $\tau \leftrightarrow \bar \tau$ 
with respect to the standard notation for the heterotic string.}. More 
precisely, one finds as expected a total derivative in moduli space.
A generic anomalous amplitude is therefore of the form 
${\cal A} = \oint_{\partial {\cal F}} d\tau 
\langle \hat V_1^{\rm unphy.} V_2^{\rm phy.} \ldots V_n^{\rm phy.} \rangle$, 
the relevant number of vertex operators being determined by the 
integration over the Grassmann fermionic zero modes. 
Importantly, such an anomaly ${\cal A}$ satisfies the Wess--Zumino (WZ)
consistency condition, and is the WZ descent of 
$I_{\cal A} = \oint_{\partial {\cal F}} d\tau \langle
V_1^{\rm phy.} V_2^{\rm phy.} \ldots V_n^{\rm phy.} \rangle$: 
${\cal A} = 2 \pi i I^{(1)}_{\cal A}$\footnote{We use here the standard
WZ descent notation. Given a gauge-invariant $D+2$ form $I$, one can define
a $D+1$ Chern-Simons form $I^{(0)}$ such that $I = d I^{(0)}$, whose 
variation is exact and defines a $D$-form $I^{(1)}$ through 
$\delta I^{(0)} = d I^{(1)}$.}.
The total anomaly polynomial $I$ encoding all the anomalous amplitudes of 
this type is then given by 
\be
I = \oint_{\partial {\cal F}} \! d\tau \, \bar A(\tau) \;,
\label{amp}
\ee
where $\bar A(\tau)$ is the generating functional of odd spin structure 
correlation functions, {i.e.} the partition function computed by deforming 
the free action with the vertex operators appearing in anomalous amplitudes. 
The result (\ref{amp}) is completely general
and its WZ descent produces the full set of anomalies. The integration 
over fermionic zero modes in $\bar A(\tau)$ automatically selects the
components of appropriate degree in any space-time dimensionality.

From a string theory point of view, the vanishing of the total anomaly 
is a consequence of modular invariance \cite{kut}--\cite{ya}. Following
\cite{kut}, one can argue that the contribution from the part of 
$\partial {\cal F}$ closing at infinity vanishes by analytic continuation 
from kinematical regions where the correlation of the factors 
$\exp i p \cdot X$ in the vertex operators give an exponential suppression. 
The rest of the integral over $\partial {\cal F}$ vanishes instead under 
the condition that the generating functional $\bar A$ is modular-invariant:
\bea
&& \bar A(\tau +1) = \bar A(\tau) \,, \label{mod1} \\
&& \bar A(-1/\tau) = \tau^2 \bar A(\tau) \;. \label{mod2}
\eea
This can be easily understood from Fig. 1. Indeed, the contributions 
from parts $I$ and $IV$ of $\partial {\cal F}$ cancel thanks to 
(\ref{mod1}), and similarly the contributions $II$ and $III$ cancel
by virtue of (\ref{mod2}). Finally, one finds therefore:
\be
I = 0 \;.
\label{I0}
\ee
At the level of low-energy effective action, it is well-known that
(\ref{I0}) appears as a cancellation between a one-loop anomaly and
a tree-level Green--Schwarz inflow. 
\begin{figure}[h]
\begin{picture}(300,250)(0,0)
\put(120,30){\epsfig{file=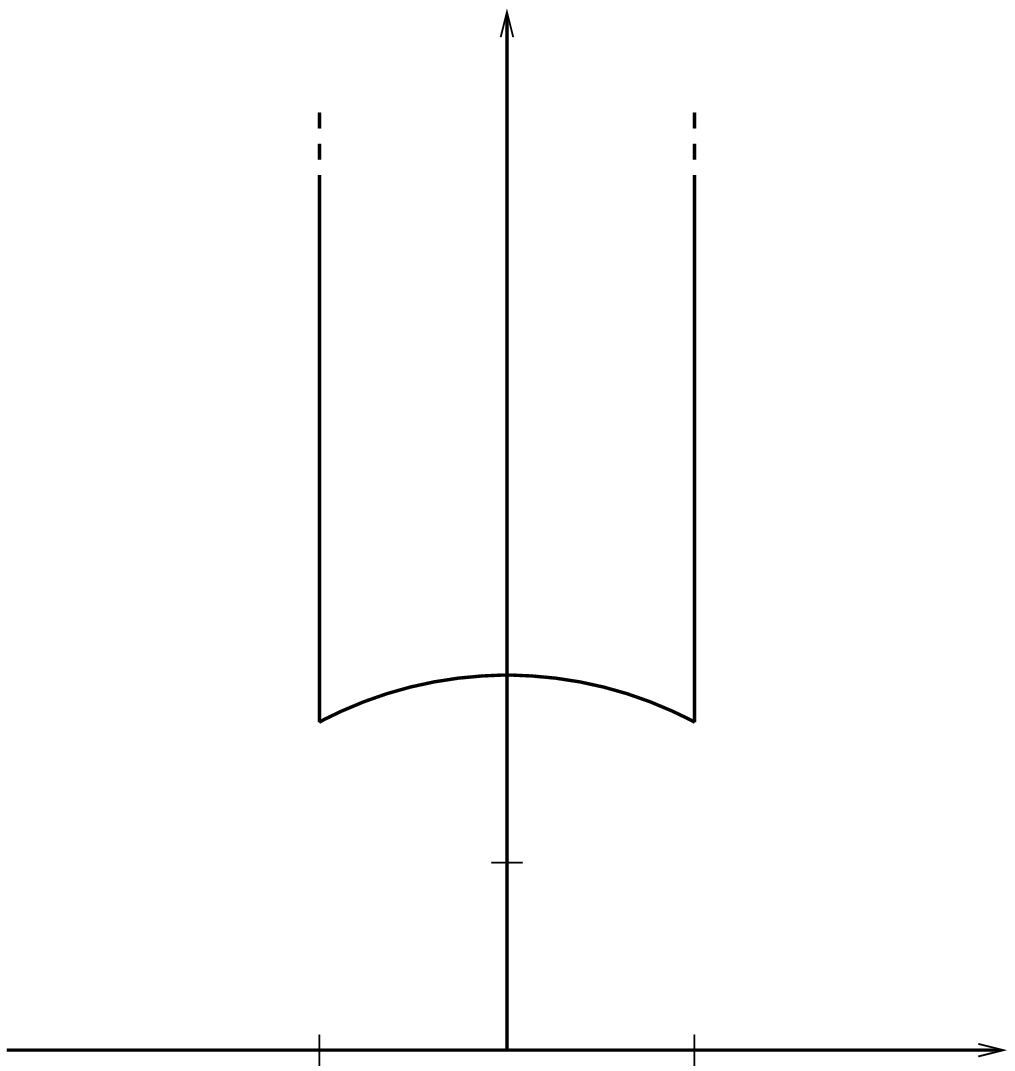,height=7cm}}
\put(190,120){${\cal F}$}
\put(168,150){$I$}
\put(253,150){$IV$}
\put(190,85){$II$}
\put(220,85){$III$}
\put(220,65){$\frac 12$}
\put(175,15){-$\frac 12$}
\put(247,15){$\frac 12$}
\put(300,15){${\rm Re}\,\tau$}
\put(225,225){${\rm Im}\,\tau$}
\end{picture}
\caption{Fundamental domain of the torus.}
\end{figure}
These two contributions can be explicitly identified in the low-energy limit 
$\alpha^\prime \rightarrow 0$ of the above string theory computation. 
In fact, it was shown in \cite{kut} that the low-energy limit of 
$\bar A(\tau)$ is technically equivalent to its $\tau \rightarrow i\infty$ 
limit. Using this result, it is possible to disentangle the contribution 
corresponding to the one-loop anomaly from the contribution of the GS inflow, 
which is seen to factorize on a pole corresponding to the exchange of the 
NSNS two-form $B$.

The low-energy interpretation can be made more concrete by explicitly
evaluating the generating functional $\bar A$ at leading order in 
$\alpha^\prime \rightarrow 0$. In this limit, the path integral 
representation can be evaluated exactly, using a modular-invariant 
regularization. It becomes clear that $\bar A$ is intimately related to 
a character-valued index $A$, obtained from the same path-integral but 
with a holomorphic regularization. This is a generalization of the 
situation described in \cite{aw}--\cite{lerche}
to the case of orbifold theories with an additional target-space 
background. With a slight abuse of language, we will simply refer to 
$A$ and $\bar A$ as elliptic genera in the following. 
On general grounds, the modular-invariant elliptic genus $\bar A$ differs 
from the holomorphic elliptic genus $A$ only by a so-called modular anomaly 
involving some four-form $X_4$ \cite{lnsw}. The precise relation is
\be
\bar A(\tau) = \exp \Bigg\{\!-\!\frac {X_4}{64 \pi^3\,{\rm Im}\tau}
\Bigg\} A(\tau) \;,
\label{AA}
\ee
where $X_4$ is in fact entirely determined by the requirement of modular 
invariance, and can be computed in a direct way using a specific 
modular-invariant regularization prescription (see appendix D). 

The appearance of the modular-invariant version $\bar A$ of the elliptic 
genus $A$ in the expression for the anomaly polynomial is of course not a 
coincidence. Indeed, it is well known that its holomorphic companion $A$ 
represents the chiral index of the full string spectrum, and the one-loop 
field-theory anomaly associated to chiral massless states is therefore given 
by \cite{aw}
\be
I_{FT} = \lim_{\tau \rightarrow i\, \infty} A(\tau) = X_6 \;,
\label{FT}
\ee
where the second equality anticipates that only the six-form 
component of the result is relevant in $D=4$.
The GS term, in turn, can be obtained by generalizing the work of 
\cite{lnsw} to the case at hand. Its expression involves the 
modular-invariant elliptic genus $\bar A$ and is given by
\be
L_{GS} =  B \wedge \Bigg[\frac 1{64 \pi^2} 
\int_{\cal F} \frac {d^2 \tau}{({\rm Im}\tau)^2} 
\bar A(\tau) \Bigg] \;,
\label{GS}
\ee
where $B$ is the NSNS two-form, dual to the axion in four dimensions. 
Using then the general expression (\ref{AA}) into (\ref{GS}), and going 
through the same manipulations as in \cite{lnsw}, the modular integral 
yielding the GS term can easily be evaluated. In $D=4$, one finds
\be
L_{GS} = - 2 \pi \, B \wedge X_2 \;,
\ee
where the two-form $X_2$ is formally defined as $X_2 = X_6/X_4$.
Clearly, this makes sense only if $X_6 = X_2 \wedge X_4$, {i.e.} only if 
the field theory anomaly factorizes. This must be guaranteed by the 
form $(\ref{AA})$ of the elliptic genus, that is modular invariance.
The induced inflow of anomaly is $I_{GS} = - X_6$ if and only if 
the Bianchi identity satisfied by the field strength $H$ of the $B$ 
field is 
\be
d H = X_4 \;.
\ee 

Summarizing, the vanishing of the total string theory anomaly is 
interpreted at low energies as the cancellation of the field-theory 
anomaly through the GS mechanism: $I_{GS} = - I_{FT}$. The generalization 
of the arguments of $\cite{susu,miki,ya,kut}$ demonstrates that anomaly 
cancellation at the string-theory level is a consequence of modular 
invariance, whereas the generalization of the arguments of \cite{lnsw,kut} 
allows us to interpret the field-theory counterpart of the cancellation 
mechanism. Intuitively, one can think of the field-theory anomaly as the $A$ 
part of $\bar A$, and attribute the GS inflow to the anomalous phase through 
which $\bar A$ differs from $A$.

As a remark, we would like to propose a very naive but suggestive
alternative way to understand how the vanishing of the anomaly is
achieved at the string-theory level. It is motivated by the unpleasant 
feature that the above arguments involve killing the contribution from 
infinity in $\partial {\cal F}$ by analytic continuation, whereas 
(\ref{FT}) suggests that this contribution should instead be directly 
linked to the field-theory anomaly. The point is that the field-theory 
Schwinger parameter associated to the modulus $\tau$ is actually
$\alpha^\prime \tau$, so that the various contributions to the anomaly 
appear from potentially different corners of moduli space, depending on 
whether one takes $\alpha^\prime \rightarrow 0$ from the beginning or 
only at the end. This is also the essence of the already mentioned 
observation of \cite{kut} about the technical equivalence of the two 
limits $\alpha^\prime\rightarrow 0$ and $\tau \rightarrow i\infty$ 
in anomalous amplitudes. In particular, if one takes brutally 
$\alpha^\prime \rightarrow 0$ from the beginning, one can no longer 
kill the contribution at infinity in $\partial {\cal F}$ by 
analytic continuation, since each pair of momenta is accompanied 
by an $\alpha^\prime$. Similarly, the term responsible for the 
factorization of the amplitude (\ref{amp}) on a $B$-pole is suppressed.
These two effects seem to compensate each other, and as a matter of fact, 
one could recover a natural interpretation of the cancellation mechanism 
by taking the $\alpha^\prime \rightarrow 0$ from the beginning and 
computing the elliptic genus with a holomorphic rather than 
modular-invariant regularization. The total anomaly (\ref{amp}) would 
then involve $A$ instead of $\bar A$, and the vanishing of the contour 
integral would be attributed to holomorphicity of $A$ in ${\cal F}$, rather 
than modular invariance. However, the relevant pieces of $\partial {\cal F}$ 
are now different. The contribution at infinity yields the field-theory 
anomaly. The contributions from $I$ and $IV$ still cancel, because 
(\ref{mod1}) is still satisfied for $A$. The contributions from $II$ and 
$III$, instead, no longer cancel since (\ref{mod2}) fails for $A$, but rather 
combine to yield the GS inflow. Although this alternative argument cannot 
be taken too seriously, we find it quite suggestive. 

\section{Orbifolds and elliptic indices}

In this section, we shall provide a concrete example of the general 
features discussed in the previous section, by studying all possible 
gauge/gravitational/target-space anomalies for $\Z_N$ 
orbifold models with standard embedding and $N$ odd, {i.e.} 
the $\Z_3$ and $\Z_7$ models. According to the arguments above, this 
involves computing the elliptic genus in a gauge, gravitational and 
target-space background, in a sense that we shall now make more precise.

To begin with, let us briefly recall some basic facts about $\Z_N$ 
heterotic orbifolds. The generator of the orbifold action is defined by
a twist vector $v_i$, where $i=1,2,3$ label the three internal 
tori and the associated complex coordinates, satisfying the condition 
$\sum_i v_i = 0$. For standard embedding, the orbifold action on the gauge
lattice is simply a shift by $v_i$ itself and acts on an $SU(3)$ part of 
the Cartan subgroup $SO(16)$ of one of the two $E_8$ factors 
($v_j = 0$ for $j > 3$). The total gauge group is then 
$E_8 \times E_6 \times H$, where $H$ is a subgroup of $SU(3)$
(see Table 1).

It is well known that these heterotic vacua possess a target-space 
duality symmetry. This symmetry acts as $SL(2,R)_i$ transformations 
on the three universal diagonal K\"ahler moduli superfields $T_i$
\cite{DVV}:
\be
T_i \rightarrow \frac {a_i T_i - i b_i}{i c_i T_i + d_i} \;.
\label{sl2r} 
\ee
Its discrete $SL(2,{\bf Z})$ subgroup is an exact symmetry (to all orders
in $\alpha^\prime$ and $g_s$) of the model, T-duality.
At tree level, (\ref{sl2r}) is a symmetry of the low-energy $N=1$ 
supergravity effective action, provided one transforms in an appropriate 
way each matter chiral superfield (and the superpotential) as well. 
It turns out that these 
transformations are always of the form:
\be
\Phi_s \rightarrow \exp \Big\{\!-\! n_i \ln (i c_i T_i + d_i)\Big\}
\Phi_s \;,
\label{matter}
\ee
with given coefficients $n_i$, the so-called modular weights \cite{il}. 
Since the transformations (\ref{sl2r}) and (\ref{matter}) act as chiral 
rotations on the component fermions, this symmetry is potentially anomalous. 
At one loop, anomalies appear through triangular graphs where the 
connection $Z_i$ associated to the target-space symmetry (\ref{sl2r}) 
appears, possibly together with gluons and gravitons, among the external 
states (see \cite{bg1}--\cite{bg2} for details). Although such anomalies 
(mainly mixed target-space/gauge and target-space/gravitational) have 
received quite a lot of attention in the past, no direct, explicit and 
complete computation of them has so far appeared in the literature. 
A satisfactory understanding of their structure is also missing. 
Since the target-space connections $Z_i$ are composite 
and not elementary fields, such anomalies are typically derived
by analogy with standard gauge anomalies associated to an elementary 
connection. This approach was adopted also in \cite{bg2}, the only 
reference considering target-space/gauge/gravitational anomalies in
full generality.

The approach we follow here is more direct: we will compute anomalous 
amplitudes involving the elementary constituents of the composite 
connection. We focus on the dependence of the target-space connection 
$Z_i$ on the diagonal untwisted moduli. These are defined as:
\be
T_i = G_{i\bar \imath} + i B_{i \bar \imath} \;,
\label{Tdef}
\ee
where the scalars $G_{i\bar \imath}$ and the pseudoscalars $B_{i\bar \imath}$
represent the internal metric and $B$-field components of the orbifold 
in complex coordinates along each of the three $T_i^2$ tori. It is 
straightforward to determine the most general form of the target-space 
connections $Z_i$ and their associated field strengths $G_i$ in terms of 
these fields. The total connection one-form is 
$Z = \sum_i (\alpha_i dT_i + \bar \alpha_{\bar \imath} d\Tb_{\bar \imath})$,
where the $\alpha_i$'s are some functions of the $T$-moduli. 
Correspondingly, the general form of the field strength $G = dZ$ is
\be
G = \frac 12 \sum_{i,j} \Big[\alpha_{i,j} \, dT_i \wedge dT_j +
\bar \alpha_{\bar \imath, \bar \jmath} \, 
d\Tb_{\bar \imath} \wedge d\Tb_{\bar \jmath} +
(\alpha_{i, \bar \jmath} - \bar \alpha_{\bar \jmath, i}) \, 
dT_{i} \wedge d\Tb_{\bar \jmath} \Big]\;.
\label{Gtot}
\ee
From (\ref{Gtot}), it is clear that the composite field strength is always
at least quadratic in the fluctuations of the moduli (\ref{Tdef}). 
Correspondingly, one can reliably compute anomalous triangular graphs with 
external composite fields by replacing each of them with a couple of 
$T$ fields. Actually, it is known from supergravity that the coefficients
$\alpha_i$ are determined by the K\"ahler potential $K$ as 
$\alpha_i = -i \partial K/\partial T_i = -i K_i$. The field strength 
(\ref{Gtot}) should therefore reduce to:
\be
G = - i \sum_{i,j} K_{i, \bar \jmath} \, dT_{i} \wedge d\Tb_{\bar \jmath} \;.
\label{Gtots}
\ee
Finally, the relevant part of the K\"ahler potential for the models under 
consideration is $K = - \sum_i\ln (T_i + \Tb_{\bar \imath})$, yielding a
diagonal result: $K_{i, \bar \jmath} = \delta_{i, \bar \jmath} / 
2 (T_i + \Tb_{\bar \imath})^2$. 
In the following, we shall assume only the general form (\ref{Gtot}), 
and verify that the supergravity result (\ref{Gtots}) is correctly 
recovered.

In order to evaluate the elliptic genus as a function of the gauge, 
gravitational and target-space backgrounds, we need to analyse the 
corresponding gluon, graviton and moduli vertex operators. 
According to the general discussion of section 2, we have 
to take into account only physical operators in the $0$-picture.
Consider first the standard case of photons and gravitons. 
Denoting with $J^a$ the current operator of the gauge current algebra,
and with $Q_a$ its zero mode, one has:
\bea
V_g &=& \xi_{\mu\nu}(p)\, \int\! d^2z\, \partial X^\mu (z) \,
\left(\bar \partial X^\nu 
+ i p \cdot \psi \psi^\mu \right)(\bar z) \,
\, e^{ip\cdot X(z,\bar z)} \;, \\
V_\gamma &=& \xi_\mu^a(p)\,\int \! d^2z \, J^a(z)\, 
\left(\bar \partial X^\mu 
+ i p \cdot \psi \psi^\mu \right)(\bar z) \,
e^{ip\cdot X(z,\bar z)} \;.
\eea
In the low-energy limit $\alpha^\prime\rightarrow 0$, one can restrict 
to the terms providing a minimal number of momenta and a maximal number 
of fermionic zero-modes, and one finds:
\bea
V^{\rm eff.}_g (R) &=& R_{\mu\nu}\, \int\! d^2z\, 
X^\mu(z)\,\partial X^\nu(z) \;, \\ 
V^{\rm eff.}_\gamma (F) &=& F^a\, Q^a\;,
\eea
where $F^a$ and $R_{\mu\nu}$ represent the gauge and gravitational curvatures
and are given by
\be
F^a = \frac 12 \, F_{\mu\nu}^a \, \psi_0^\mu \psi_0^\nu \;,\;\;
R_{\mu\nu} = \frac 12\, R_{\mu\nu\rho\sigma}\, \psi_0^\rho \psi_0^\sigma \;.
\label{curvgg2}
\ee
Consider next the case of the $T_i$ moduli. Their vertex operators 
are given by\footnote{In (\ref{VTb}), $\bar p$ denotes simply the 
momentum of $\bar T$, not the complex conjugate of $p$.}:
\bea
V_{T_i} &=& T_i(p)\int\! d^2z \, \partial \Xb^{\bar \imath}(z)
\Big(\bar \partial X^i + i p \cdot \psi \psi^i\Big)(\bar z)\,
e^{ip\cdot X(z,\bar z)} \;, \label{VT} \\
V_{\Tb_{\bar \imath}} &=& \Tb_{\bar \imath} (\bar p)
\int\! d^2z \, \partial X^i(z)
\Big(\bar \partial \Xb^{\bar \imath} + 
i \bar p \cdot \psi \psib^{\bar \imath}\Big)(\bar z)\,
e^{i\bar p\cdot X(z,\bar z)} \;, \label{VTb}
\eea
and lead to:
\bea
&& V_{T_i}^{\rm eff.} = d T_i \int\! d^2z \, 
\psi^i (\bar z)\, \partial \Xb^{\bar \imath}(z) \;,\label{Teff} \\
&& V_{\Tb_{\bar \imath}}^{\rm eff.} =  d \Tb_{\bar \imath} \int\! d^2z \, 
\psib^{\bar \imath}(\bar z)\, \partial X^i(z) \;,
\label{Tbeff}
\eea
where
\be
d T_i = i p_\mu T_i \, \psi_0^\mu \;,\;\; 
d \Tb_{\bar \imath} = i \bar p_\mu \Tb_{\bar \imath} \, \psi_0^\mu \;.
\ee
It is clear from (\ref{Teff}) and (\ref{Tbeff}) that the only non-vanishing 
contractions among moduli occur between a given $T_i$ modulus and its 
complex conjugate $\Tb_{\bar \imath}$. A generic correlation is 
therefore non-vanishing only if it includes an equal number of 
$T_i$ and $\Tb_{\bar \imath}$ vertices.
This general property can be easily understood also in the 
path-integral representation of the generating functional. Indeed, 
integrating out the internal fermions appearing in (\ref{Teff}) and 
rescaling the internal bosonic fields to normalize their kinetic terms, 
one ends up with an effective bosonic interaction that is interpreted as 
the effective vertex operator for the composite connection. This 
manipulation is equivalent to grouping in all possible ways the moduli 
vertices in $T_i$--$\Tb_{\bar \imath}$ pairs and explicitly performing the 
fermionic contraction within each pair. One finds:
\be 
V_{Z_i}^{\rm eff.}(G_i)= G_i \int\! d^2z \, \Xb^{\bar \imath} (z) \, 
\partial X^i (z) \;,
\ee
with $G_i$ representing the target-space curvature two-forms and given by
\be
G_i = \frac i{2(T_i+\Tb_{\bar \imath})^2} \, p_\mu \, T_{i} \, 
\bar p_\nu \,
\Tb_{\bar \imath} \, \psi_0^\mu \, \psi_0^\nu \;.
\ee

As expected, the external moduli automatically group in 
$T_i$--$\Tb_{\bar \imath}$ pairs, reconstructing composite 
connections of the form (\ref{Gtots}) with the correct K\"ahler potential. 
Correpondingly, one can define three independent connections and 
curvatures associated to each internal torus and given by
\bea
&& Z_i = \frac {-i}{(T_i+\Tb_{\bar \imath})} 
\, d(T_{i} - \Tb_{\bar \imath}) \;, 
\label{Zi} \\
&& G_i = \frac {-i}{2(T_i+\Tb_{\bar \imath})^2} 
\, dT_{i} \wedge d\Tb_{\bar \imath} \;.
\label{Gi}
\eea
It is important to stress that within this direct approach, the pairing 
of the moduli appearing as external states into composite connections 
is a derived property and not an assumption. This mechanism is depicted
in Fig. 2 for a generic anomalous amplitude.

\begin{figure}[h]
\begin{picture}(300,210)(0,0)
\put(140,25){\epsfig{file=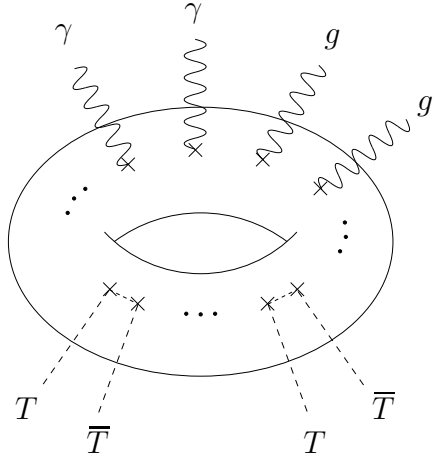,height=5cm}}
\put(158,167){$\gamma$}
\put(207,175){$\gamma$}
\put(260,165){$g$}
\put(295,140){$g$}
\put(143,23){$T$}
\put(170,10){$\Tb$}
\put(252,10){$T$}
\put(278,23){$\Tb$}
\end{picture}
\caption{Pairing of moduli in anomalous amplitudes.}
\end{figure}

The computation of the elliptic genus does not present any problem, since
all the effective vertices are quadratic. We begin with the holomorphic 
regularization. The result factorizes into space-time, compact and gauge 
partition functions:
\be
A(\tau;F,R,G) = \frac 1{2N} \sum_{k,l=0}^{N-1} N_{k,l} \,  
Z_{ST}(R) \, Z_{C}^{k,l}(G) \, Z_{\Gamma}^{k,l}(F) \;.
\ee
The $1/2N$ factor is due to the GSO projection in a $\Z_N$ orbifold model
and $N_{k,l}$ represents the number of points that are at the same time
$k$- and $l$-fixed. Restricting to the six-form component of the elliptic genus
relevant in four dimensions, one can rescale the free-particle normalization
to cancel the $\tau$-dependence coming from the curvatures. The only $\tau$ 
dependence that is left over arises then through $q = e^{2 i \pi \tau}$.
For $(k,l) \neq (0,0)$, one finds the following results in terms of the 
skew eigenvalues $y_i$, $\lambda_a$ and $g_i$ of the rescaled 
gauge, gravitational and target space curvatures (see appendix A)
(recall that $v_j = 0$ for $j>3$)\footnote{The definition of twisted 
$\theta$-functions used here is taken from appendix A of \cite{ss4}.}:
\bea
&& Z_{ST}(R) = \eta^2(\tau) \prod_{a=1}^2 \frac{(i \lambda_a)\eta (\tau)}
{\theta_1(i\lambda_a/2\pi|\tau)} \;, \label{ZTot1} \\
&& Z_{C}^{k,l}(G) = \prod_{i=1}^3 \frac{\eta (\tau)}
{\theta_1{l v_i \brack kv_i}(-g_i/2\pi|\tau)} \;, \raisebox{23pt}{}
\label{ZTot2} \\
&& Z_\Gamma^{k,l}(F) = \frac 14 \sum_{\alpha,\beta} \prod_{i=1}^8
\frac{\theta_\alpha{l v_i \brack kv_i}(- y_i/2 \pi|\tau)}{\eta(\tau)} 
\prod_{h=1}^8 \frac{\theta_\beta(- y^\prime_h/2 \pi|\tau)}{\eta(\tau)} \;.
\hspace{30pt} \label{ZTot3}
\eea
In the completely untwisted sector $(k,l)=(0,0)$, there are six additional 
fermionic zero modes arising from the compact dimensions, and the result 
is apparently vanishing. Actually, there is a potential contribution 
from this sector to pure target-space anomalies of the form $G_1G_2G_3$. 
This can be easily understood by noting from (\ref{Teff}) and (\ref{Tbeff}) 
that in this case it is possible to soak the internal fermionic zero modes 
as well. However, the correlation that one obtains involves only internal 
right-moving momenta, and is clearly irrelevant in the limit (\ref{FT}) 
and for the exponential term in (\ref{AA}). Hence, it will be completely 
neglected in the following\footnote{From a more physical point of view,
it is clear that the completely untwisted $(0,0)$ sector, having $N=4$ 
supersymmetry, cannot give rise to any anomaly.}.

The corresponding result in the modular-invariant regularization summarized 
in appendix D is found to have the expected form (\ref{AA}) with 
\be
X_4(F,R,G) = 8 \pi^2 \Big(\sum_{a=1}^2 \lambda_a^2 + \sum_{i=1}^3 g_i^2 
- \sum_{i=1}^8 y_i^2 - \sum_{h=1}^8 {y_h^\prime}^2 \Big) \;.
\label{an}
\ee
This is just what is needed for $\bar A$ to satisfy the modular 
transformation properties (\ref{mod1}) and (\ref{mod2}). 
Indeed, it is not difficult, although a little bit laborious,
to verify that:
\bea
\bar A(\tau +1;F,R,G) &=& \bar A (\tau;F,R,G) \;, \nn \\
\bar A(-1/\tau; F,R,G) &=& \frac 1\tau \bar A (\tau; F\tau,R\tau,G\tau)
= \tau^2 \bar A (\tau;F,R,G) \;,
\label{modinv}
\eea
where the last step in the second equation holds for the relevant 
six-form component.

Before entering into the details of the expansion of eqs. 
(\ref{ZTot1})--(\ref{ZTot3}), it is convenient to define the quantity
\be
C_k = \prod_{i=1}^3 \sin \pi k v_i \;,
\ee
and the representative
\be
\theta_i^l = l v_i - {\rm int}(l v_i) 
\ee
of the generic twist vector $l v_i$ in the interval $[0,1]$. 
Whereas $\sum_i v_i$ is $0$ in order to have $N=1$ supersymmetry, 
$\sum_i \theta_i^l$ is $1$ for half of the twisted sectors and $2$ 
for the conjugate half. More precisely, one can define the sign 
$\epsilon_l$ distinguishing between conjugate twisted sectors as
\be
\epsilon_l = -(-1)^{\sum_i \theta_i^l} = - {\rm sign}(C_l) \;.
\ee
We also define $S_l$ to be the set of twisted sectors where $\epsilon_l=1$.
The set $S_l$ is $\{1\}$ for $\Z_3$ and $\{1,2,4\}$ for $\Z_7$.

\subsection{Field theory anomaly}

The field theory anomaly $I_{FT}$ can now be computed according to (\ref{FT}). 
Taking the limit $\tau \rightarrow i\, \infty$ of $A$ turns out to
be a rather non-trivial exercise. For this reason, we present here only the 
final result, leaving the relevant details to appendices B and C. As expected,
simple characteristic classes, defined in appendix A, are reconstructed.
Since the visible gauge group is $E_6 \times H$, with $H \subset SU(3)$ 
depending on the orbifold projection, it is useful to refer to the underlying 
$E_6 \times SU(3)$ group common to all orbifold models with standard 
embedding. In the untwisted sector, one finds
\bea
&& I_{FT}^{\rm untw.}(F,R,G) = \frac 1{2N} \sum_{k=1}^{N-1} C_k \,
\Bigg[\widehat A_k(G) \, \widehat G(R) 
+ \widehat G_k(G) \, \widehat A(R) \nn \\
&& \hspace{155pt} + \Big({\rm ch}_{\bf 248}(F)
+ {\rm ch}_{({\bf 78},{\bf 1})}(F) + {\rm ch}^k_{({\bf 1},{\bf 8})}(F) \nn \\
&& \hspace{170pt} +\, {\rm ch}^k_{({\bf 27},{\bf 3})}(F)
+ {\rm ch}^k_{({\bf \overline{27}},{\bf \overline{3}})}(F)\Big) 
\widehat A_k(G) \, \widehat A(R) \Bigg] \;, \raisebox{20pt}{}
\hspace{30pt} 
\label{untwisted}
\eea
in terms of the twisted Chern characters ${\rm ch}^k_{\rho}(F)$
defined in (\ref{chk}), (\ref{chq}). In the twisted sectors, finds instead 
the following compact expression for the anomaly:
\bea
&& I_{FT}^{\rm tw.}(F,R,G) = - i \sum_{l\in S_l} N_{l} \,
\Bigg[{\rm ch}_{({\bf 27},{\bf 1})}(F)\, 
{\rm ch}_{\theta_i^l - \frac 13}(F)\,
{\rm ch}_{-(1-\theta_i^l)}(G) \nn \\
&& \hspace{135pt} + \sum_{s}\,{\rm ch}_{q_i(s)}(F) \,
{\rm ch}_{n_i(s)}(G) \Bigg] \, \widehat A(R) \;,
\label{twisted}
\eea
where the last sum runs over all the massless $E_6$-neutral twisted 
states $s$, with modular weights $n_i(s)$ and $SU(3)$ charge vector 
$q_i(s)$. The structure of these states is described in
more detail in appendix C. Their contributions can be read off directly 
from the elliptic genus, and the correct modular weights and charge vectors 
are found. The Chern characters ${\rm ch}_{q_i}(F)$ and ${\rm ch}_{n_i}(G)$ 
are defined in (\ref{chq}) and (\ref{chn}), and the complete massless 
spectra are reported in Table 1.

\begin{table}[h]
\vbox{
$$\vbox{\offinterlineskip
\hrule height 1.1pt
\halign{&\vrule width 1.1pt#
&\strut\quad#\hfil\quad&
\vrule#
&\strut\quad#\hfil\quad&
\vrule#
&\strut\quad#\hfil\quad&
\vrule#
&\strut\quad#\hfil\quad&
\vrule#
&\strut\quad#\hfil\quad&
#
&\strut\quad#\hfil\quad&
\vrule width 1.1pt#\cr
height3pt
&\omit&
&\omit&
&\omit&
&\omit&
&\omit&
&\omit&
\cr
&\hfil $P$&
&\hfil $v$&
&\hfil $\!n_T\!$&
&\hfil $H$&
&\hfil&
&\hspace{-40pt} Matter&
\cr
height3pt
&\omit&
&\omit&
&\omit&
&\omit&
&\omit&
&\omit&
\cr
\noalign{\hrule height 1.1pt}
height3pt
&\omit&
&\omit&
&\omit&
&\omit&
&\omit&\vline
&\omit&
\cr
&$\eqalign{{\;\,}\Z_3 \\ \\}$&
&$\eqalign{(\frac 13, \frac 13, \frac {-2}3) \\ \\}$&
&$\eqalign{{\;\;\,}9 \\ \\}$&
&$\eqalign{SU(3) \\ \\}$&
&$\eqalign{1\;\,: ({\bf 27},{\bf 3})_{-\delta_i^{1,2,3}} \\ \\}$&\vline
&$\eqalign{g\;: ({\bf 27},{\bf 1})_{-\rho_i^{1,2,3}} \\ \;\;\;\;\;\; 
({\bf 1},{\bf \overline{3}})_{-(\rho_i^1+\delta_i^{1,2,3})} \\}$&
\cr
height3pt
&\omit&
&\omit&
&\omit&
&\omit&
&\omit&\vline
&\omit&
\cr
\noalign{\hrule}
height3pt
&\omit&
&\omit&
&\omit&
&\omit&
&\omit&\vline
&\omit&
\cr
&$\eqalign{{\;\,}\Z_7 \\ \\ \\ \\ \\ \\ \\
\\ \\ \\}$&
&$\eqalign{(\frac 17,\frac 27,\frac {-3}7)
\\ \\ \\ \\ \\ \\ \\
\\ \\ \\}$&
&$\eqalign{{\;\;\,}3 \\ \\ \\ \\ \\ \\ \\
\\ \\ \\}$&
&$\eqalign{U(1)^2 \\ \\ \\ \\ \\ \\ \\
\\ \\ \\}$&
&$\eqalign{
1\;\,: ({\bf 27}_{0,2})_{-\delta_i^1} \\ \;\;\;\;\;\;\,
({\bf 27}_{1,\mbox{-}1})_{-\delta_i^2} \raisebox{13.5pt}{} \\ \;\;\;\;\;\;\,
({\bf 27}_{1,3})_{-\delta_i^3} \raisebox{13.5pt}{} \\ \;\;\;\;\;\;\, 
({\bf 1}_{1,\mbox{-}3})_{-\delta_i^1} \raisebox{13.5pt}{} \\ \;\;\;\;\;\;\,
({\bf 1}_{\mbox{-}2,0})_{-\delta_i^2} \raisebox{13.5pt}{} \\ \;\;\;\;\;\;\,
({\bf 1}_{\mbox{-}1,\mbox{-}1})_{-\delta_i^3}\raisebox{13.5pt}{} 
\\ \vspace{38pt}}$& \vline
&$\eqalign{
g\;: ({\bf 27}_{\frac {\mm 2}7, \frac {\mm 4}7})_{-\rho_i^1} \\
\;\;\;\;\;\; 
({\bf 1}_{\frac {\mm 2}7, \frac {\mm 18}7})_{
-(\rho_i^1 + \delta_i^{1})} \\ \;\;\;\;\;\; 
({\bf 1}_{\frac {\mm 9}7, \frac {3}7})_{
-(\rho_i^1 + \delta_i^{2})} \\ \;\;\;\;\;\; 
({\bf 1}_{\frac {5}7, \frac {3}7})_{
-(\rho_i^1 + \delta_i^{3})} \\ \;\;\;\;\;\; 
({\bf 1}_{\frac {\mm 9}7, \frac 37})_{
-(\rho_i^1 + 2 \delta_i^{1})} \\ \;\;\;\;\;\; 
({\bf 1}_{\frac 57, \frac 37})_{
-(\rho_i^1 + 2 \delta_i^{2})} \\ \;\;\;\;\;\; 
({\bf 1}_{\frac 57, \frac 37})_{
-(\rho_i^1 + \delta_i^1 - \delta_i^3)} \\ \;\;\;\;\;\; 
({\bf 1}_{\frac 57, \frac 37})_{
-(\rho_i^1 + 2 \delta_i^1 + \delta_i^2)} \\ \;\;\;\;\;\; 
({\bf 1}_{\frac 57, \frac 37})_{
-(\rho_i^1 + 4 \delta_i^1)}}$&
\cr
height3pt
&\omit&
&\omit&
&\omit&
&\omit&
&\omit&\vline
&\omit&
\cr
\cline{10-12}
height3pt
&\omit&
&\omit&
&\omit&
&\omit&
&\omit&\vline
&\omit&
\cr
&\omit& 
&\omit&
&\omit&
&\omit&
&$\eqalign{
g^2: ({\bf 27}_{\frac 37, \frac {\mm 1}7})_{-\rho_i^2} \\ 
\;\;\;\;\;\;
({\bf 1}_{\frac {10}7, \frac 67})_{
-(\rho_i^2 + \delta_i^{3})} \\ \;\;\;\;\;\; 
({\bf 1}_{\frac 37, \frac {\mm 15}7})_{
-(\rho_i^2 + \delta_i^{1})} \\ \;\;\;\;\;\; 
({\bf 1}_{\frac {\mm 4}7, \frac 67})_{
-(\rho_i^2 + \delta_i^{2})} \\ \;\;\;\;\;\; 
({\bf 1}_{\frac 37, \frac {\mm 15}7})_{
-(\rho_i^2 + 2 \delta_i^{3})} \\ \;\;\;\;\;\; 
({\bf 1}_{\frac {\mm 4}7, \frac 67})_{
-(\rho_i^2 + 2 \delta_i^{1})} \\ \;\;\;\;\;\; 
({\bf 1}_{\frac {\mm 4}7, \frac 67})_{
-(\rho_i^2 + \delta_i^3 - \delta_i^2)} \\ \;\;\;\;\;\; 
({\bf 1}_{\frac {\mm 4}7, \frac 67})_{
-(\rho_i^2 + 2 \delta_i^3 + \delta_i^1)} \\ \;\;\;\;\;\; 
({\bf 1}_{\frac {\mm 4}7, \frac 67})_{
-(\rho_i^2 + 4 \delta_i^3)}}$&\vline
&$\eqalign{
g^4: ({\bf 27}_{\frac {\mm 1}7, \frac 57})_{-\rho_i^4} 
\raisebox{18pt}{} \\ \;\;\;\;\;\;
({\bf 1}_{\frac {\mm 8}7, \frac {12}7})_{
-(\rho_i^4 + \delta_i^{2})} \\ \;\;\;\;\;\; 
({\bf 1}_{\frac 67, \frac {12}7})_{
-(\rho_i^4 + \delta_i^{3})} \\ \;\;\;\;\;\; 
({\bf 1}_{\frac {\mm 1}7, \frac {\mm 9}7})_{
-(\rho_i^4 + \delta_i^{1})} \\ \;\;\;\;\;\; 
({\bf 1}_{\frac 67, \frac {12}7})_{
-(\rho_i^4 + 2 \delta_i^{2})} \\ \;\;\;\;\;\; 
({\bf 1}_{\frac {\mm 1}7, \frac {\mm 9}7})_{
-(\rho_i^4 + 2 \delta_i^{3})} \\ \;\;\;\;\;\; 
({\bf 1}_{\frac {\mm 1}7, \frac {\mm 9}7})_{
-(\rho_i^4 + \delta_i^2 - \delta^i_1)} \\ \;\;\;\;\;\; 
({\bf 1}_{\frac {\mm 1}7, \frac {\mm 9}7})_{
-(\rho_i^4 + 2 \delta_i^2 + \delta^i_3)} \\ \;\;\;\;\;\; 
({\bf 1}_{\frac {\mm 1}7, \frac {\mm 9}7})_{
-(\rho_i^4 + 4 \delta_i^2)} \vspace{5pt}}$&
\cr 
height3pt 
&\omit& 
&\omit&
&\omit&
&\omit&
&\omit&\vline
&\omit&
\cr
}
\hrule height 1.1pt}
$$
\caption{The $\Z_3$ and $\Z_7$ models, with their twist $v$, 
enhancement gauge group $H$, number $n_T$ of K\"ahler moduli and 
matter spectrum. The representations refer to the visible gauge group 
$G=E_6 \times H$ and have multiplicity equal to the number of fixed points 
in each twisted sector. The modular weights $n_i$ of each matter field appear
as a subindex ($\rho^l_i = 1 - \theta_i^l$).}}
\end{table}

As discussed in last section, consistency requires that the total anomaly 
must take a factorized form, as a consequence of modular invariance.
We will now verifiy this explicitly for the two models under consideration.
Irreducible gauge-anomalies cancel between untwisted an twisted sectors.
Similarly, the structure of the cubic target-space anomalies drastically 
simplifies in the total results, in such a way to allow a simple factorization
compatible with modular invariance.

\vskip 9pt
\noindent
{\bf $\Z_3$ model}
\vskip 3pt
\noindent
For the $\Z_3$ model, $H=SU(3)$. Consider first the untwisted sector. 
The twisted Chern characters appearing in (\ref{untwisted}) are easily 
written in terms of the ordinary ones. Defining $\alpha = \exp 2 \pi i/3$, 
one finds\footnote{We have that $\alpha^k {\rm ch}_{R}(F)$ and 
$\alpha^{-k} {\rm ch}_{R}(-F)$ give the same contribution when 
summing over $k$.}:
\bea
&& I_{FT}^{\rm untw.} = \frac 16 \sum_{k=1}^{2} C_k \,
\Bigg[\widehat A_k(G) \, \widehat G(R) + \widehat G_k(G) \, \widehat A(R) 
\nn \\ && \hspace{93pt} +\, \Big({\rm ch}_{\bf 248}(F)
+ {\rm ch}_{({\bf 78},{\bf 1})}(F) + {\rm ch}_{({\bf 1},{\bf 8})}(F) 
\nn \\ && \hspace{109pt} +\,2\, \alpha^k {\rm ch}_{({\bf 27},{\bf 3})}(F)
\Big)\, \widehat A_k(G) \, \widehat A(R) \Bigg] \;. \raisebox{20pt}{} 
\hspace{20pt}
\label{unz3}
\eea
Consider next the twisted sector. 
As can be seen from Table 1, there is a triplet of non-oscillator states,
and from (\ref{twisted}) one gets:
\bea
&& I_{FT}^{\rm tw.}(F,R,G) = - 27\,i \, \Bigg[
{\rm ch}_{({\bf 27},{\bf 1})}(F) \, {\rm ch}_{-(1-\theta_i^1)}(G) 
\nn \\ && \hspace{114pt} + \sum_{j=1}^3
{\rm ch}_{({\bf 1},{\bf \overline{3}})}(F) 
\, {\rm ch}_{-(1 - \theta_i^1 + \delta^j_i)}(G) \Bigg]\, 
\widehat A(R) \;. \hspace{20pt}
\label{tz3}
\eea
It is then straightforward to compute the explicit expression of the total
anomaly. One finds as expected a factorized expression:
\bea
&& I_{FT} = \frac {15}{2(2 \pi)^3} \Big(\sum_{i=1}^3 G_i \Big)
\Big({\rm tr} R^2 - {\rm tr}_{E_8} F^2 - \frac 1{3}\,{\rm tr}_{E_6} F^2 
- 2 \,{\rm tr}_{SU(3)} F^2 + 2 \sum_{i=1}^3 G_i^2 \Big). \hspace{20pt}
\label{I3}
\eea
We have used the standard definition 
${\rm Tr}_{E_8} F^2  = 30\, {\rm tr}_{E_8} F^2$ for the ${\bf 248}$ of 
$E_8$ and the relation ${\rm Tr}_{E_6} F^2 = 4 \,{\rm tr}_{E_6} F^2$
between the ${\bf 78}$ and ${\bf 27}$ of $E_6$.

\vskip 9pt
\noindent
{\bf $\Z_7$ model}
\vskip 3pt
\noindent
For the $\Z_7$ model, $H=U(1) \times U(1)$. The definition of the embedding of 
$U(1) \times U(1) \subset SU(3)$ is completely arbitrary, and for convenience 
we will not yet make any precise choice. In fact, together with the 
$U(1) \subset E_6$, there are three $U(1)_i$, $i=1,2,3$, representing the 
unbroken part of the $SO(6)$ group of internal rotations. There is then 
a natural choice for this subgroup, in which $U(1)_i$ corresponds to 
rotations within the $i$-th internal torus $T^2_i$. The $i$-th component 
of the charge vector $q_i$, introduced above for a generic state contributing 
to the anomaly, then represents the charge with respect to this $U(1)_i$.
One can therefore keep the charge vector $q_i$ to describe the $H$ quantum 
numbers; since $q_i$ satisfies $\sum_i q_i = 0$, it is always in $SU(3)$
and never provides additional charge under $U(1) \subset E_6$. We will 
adopt this notation below, as was also done in \cite{dkl2}.

Consider first the untwisted sector. Again, the twisted Chern characters are 
easily reduced to the ordinary ones. In particular, one finds three 
replicas of matter contributions, which it will prove convenient to label
with $l=1,2,4$ in the same way as the three twisted sectors, and define
the corresponding permutation degree $d_l$ such that $d_1=0$, 
$d_2=1$, $d_4=2$. Defining then $\alpha = \exp 2 \pi i/7$, one finds:
\bea
&& I_{FT}^{\rm untw.} = \frac 1{14} \sum_{k=1}^{6} C_k \,
\Bigg[\widehat A_k(G) \, \widehat G(R) + \widehat G_k(G) \, \widehat A(R) 
\nn \\ && \hspace{99pt} +\, \Big({\rm ch}_{\bf 248}(F)
+ {\rm ch}_{({\bf 78},{\bf 1})}(F) + 2 \label{unz7} \\ 
&& \hspace{114pt} +\,2\!\sum_{l=1,2,4}\! \alpha^{lk} 
({\rm ch}_{{\bf 27}_{\scriptscriptstyle{
-\delta^{1+\dl}_i + \delta^{2+\dl}_i}}}(F)
+ {\rm ch}_{{\bf 1}_{\scriptscriptstyle{\delta^{\dl}_i}}}(F))
\raisebox{20pt}{}\Big)\,\widehat A_k(G)\,\widehat A(R) \Bigg] \;. \nn
\eea
Consider next the twisted sectors. The three sectors $l=1,2,4$ are related
by permutations, and in each of them there are eight massless states, 
neutral under $E_6$, whose modular weights and charges (with a specific 
choice of $U(1)$ charges, see below) are reported in Table 1.
The form of the anomaly then is
\bea
&& I_{FT}^{\rm tw.}(F,R,G) = - 7\,i \! \sum_{l=1,2,4} \Bigg[
{\rm ch}_{{\bf 27}_{\scriptscriptstyle{\theta_i - \frac 13}}}
(F)\, {\rm ch}_{-(1-\theta_i)}(G) \nn \\ && 
\hspace{135pt} + \!\sum_{j=1,2,3}\! 
{\rm ch}_{{\bf 1}_{\scriptscriptstyle{\theta_i - \delta^{j\m \dl}_i}}}(F)\, 
{\rm ch}_{-(1-\theta_i + \delta^{j\m \dl}_i)}(G) \raisebox{14pt}{} \nn \\ && 
\hspace{135pt} + \!\sum_{h=1,2}\! 
{\rm ch}_{{\bf 1}_{\scriptscriptstyle{\theta_i - \delta^{h\m \dl + 1}_i}}}(F)\,
{\rm ch}_{-(1 - \theta_i + 2 \delta^{h\m \dl}_i)}(G) 
\raisebox{13pt}{} \nn \\ && \hspace{135pt} 
+\, {\rm ch}_{{\bf 1}_{\scriptscriptstyle{
\theta_i - \delta^{3\m \dl}_i}}}(F)\, 
{\rm ch}_{-(1 - \theta_i + 4 \delta^{3\m \dl}_i)}(G) 
\raisebox{13pt}{}\nn \\ && \hspace{135pt} 
+\, {\rm ch}_{{\bf 1}_{\scriptscriptstyle{
\theta_i - \delta^{3\m \dl}_i}}}(F)\, 
{\rm ch}_{-(1 - \theta_i + 2 \delta^{1-\dl}_i + \delta^{2-\dl}_i)}(G) 
\raisebox{16pt}{} \nn \\ && \hspace{135pt} +\, 
{\rm ch}_{{\bf 1}_{\scriptscriptstyle{\theta_i - \delta^{3\m \dl}_i}}}(F)\, 
{\rm ch}_{-(1 - \theta_i + \delta^{1\m \dl}_i - \delta^{3\m \dl}_i)}(G) 
\Bigg]\, \widehat A(R) \;.\raisebox{16pt}{} \hspace{30pt} 
\label{tz7} 
\eea
In order to compute the explicit form of (\ref{tz7}), one has to choose 
a specific embedding of the $U(1)$'s. One can define them through the 
corresponding charges $Q_{1,2}$ assigned to the generic $SU(3)$ vector 
$q_i$. We take the combinations $Q_1(q) = (q_2 - q_3)/2\sqrt{2}$ and 
$Q_2(q) = (2 q_1 - q_2 - q_3)/2\sqrt{6}$, which are canonically normalized. 
Notice that the (non-canonically normalized) charge $Q_0$ with respect to 
the $U(1) \subset E_6$ is given by $Q_0(q) = 2(q_1 + q_2 + q_3)$ , which 
vanishes for any $q_i$ occurring here. The charges obtained in this way 
for each state are reported in Table 1, in units of $1/2\sqrt{2}$ for 
$Q_1$ and $1/2\sqrt{6}$ for $Q_2$. After a straightforward but lengthy 
computation, one finds:
\bea
I_{FT} &=& \frac {15}{2(2 \pi)^3} \Big(\sum_{i=1}^3 G_i \Big)
\Big({\rm tr} R^2 - {\rm tr}_{E_8} F^2  
- \frac 13 \,{\rm tr}_{E_6} F^2 
- \frac 12 \, F_1^2 - \frac 12 \, F_2^2 
+ 2 \sum_{i=1}^3 G_i^2 \Big) \;. \hspace{30pt}
\label{I7}
\eea
Notice that all $U(1)$ anomalies cancel.

\subsection{GS terms}

The GS term can be computed from (\ref{GS}), as described in last section. 
To do so, one has first of all to evaluate (\ref{an}) in each model: the
GS term is then unambiguously determined. One finds the following results
for the two models under consideration:
\bea
\Z_3 &\,:\;& X_4 = {\rm tr} R^2 - {\rm tr}_{E_8} F^2 
- \frac 1{3}\,{\rm tr}_{E_6} F^2 - 2 \,{\rm tr}_{SU(3)} F^2 
+ 2 \sum_{i=1}^3 G_i^2 \;, \label{43} \\
\Z_7 &\,:\;& X_4 = {\rm tr} R^2 - {\rm tr}_{E_8} F^2  
- \frac 13 \,{\rm tr}_{E_6} F^2 - \frac 12 \, F_1^2 
- \frac 12 \, F_2^2 + 2 \sum_{i=1}^3 G_i^2 \;.
\label{47}
\eea
The gauge part of (\ref{43}) and (\ref{47}) is universal, {i.e.} 
independent of the gauge 
group factor, as required for a GS mechanism involving only the
NSNS axion \cite{dfkz}. This can be easily verified by using the 
values $C(E_8)=30$, $C(E_6)=12$ and $C(SU(3))=3$ for the quadratic Casimirs 
to evaluate gauge traces; one finds 
${\rm tr}_{E_8} F^2 
= 1/2 \,{F_{E_8}}^n \wedge {F_{E_8}}_n$,
$1/3\,{\rm tr}_{E_6} F^2 
= 1/2 \, {F_{E_6}}^n \wedge {F_{E_6}}_n$,
$2 {\rm tr}_{SU(3)} F^2 
= 1/2 \, {F_{SU(3)}}^n \wedge {F_{SU(3)}}_n$,
where the index $n$ is summed over all the generators of the particular group 
factor.

Notice that the explicit form of $X_4$ in both models requires a very special 
factorization of the one-loop anomalies (\ref{I3}) and (\ref{I7}). 
For instance, cubic $G_1G_2G_3$ anomalies have to vanish, as indeed happens in
both models, but thanks to a very non-trivial compensation of the contributions
of all fields.

\section{Low-energy interpretation}

We shall now analyse the results of the string theory computation 
of last section from a low-energy supergravity point of view.
In particular, we will attempt to make contact with the results of 
\cite{dfkz}--\cite{bg2}.

A first important issue is the form of the anomalies derived within 
string theory, eqs. (\ref{untwisted}) and (\ref{twisted}). 
As we shall now discuss, the results that we obtain
do not seem to match in all details the expectations based on a
low-energy supergravity approach. A number of important 
qualitative differences can be understood and recognized to 
correspond to the different frames used in the two contexts: the Einstein
and the string frames. The dependence on the gauge and gravitational 
curvatures is encoded in the standard characteristic classes, as expected. 
As for the dependence on the target-space curvature, on the other hand, there 
is an important qualitative difference between the contributions of untwisted 
and twisted states; for the former, this dependence resembles a 
gravitational dependence, whereas for the latter it is more similar to a 
gauge dependence. This suggest that twisted states feel the target-space 
background only through the target-space connection 
$Z_\mu = -i \sum_i (T_i + \Tb_{\bar \imath})^{-1} 
\partial_\mu (T_i - \Tb_{\bar \imath})$, 
whereas untwisted states have some additional sensitivity to it. 
In fact, this is recognized to be nothing but the effect of the 
moduli-dependent compactification volume 
$V = \prod_i (T_i + \Tb_{\bar \imath})$,
arising from the determinant of the internal metric. In the string frame,
such a volume factor arises by dimensional reduction in the untwisted sector,
but is clearly absent in twisted sectors.
From a purely low-energy supergravity point of view, in turn, it is possible 
and in fact convenient to move to the so-called Einstein frame, where in 
particular the volume dependence in the untwisted sector part of the
effective action is reabsorbed through a suitable definition of the metric.
In this framework, it is then plausible to expect that target-space anomalies 
can be computed as simple gauge anomalies. This is indeed the approach 
adopted in \cite{bg2}. Clearly, the final result for the total anomaly should 
be identical in the two approaches, 
and at most, one expects a reshuffling of the single contributions to 
the anomaly from those states affected by the frame redefinition. 
In fact, the comparison between the string-derived anomaly encoded in 
(\ref{untwisted})-(\ref{twisted}) and that expected from an Einstein 
frame supergravity approach as in \cite{bg2} can be done, and seems to 
lead to a discrepancy.

Consider first mixed target-space-gauge and target-space-gravitational 
anomalies. The relevant components of the polynomial $I_{FT}$ given by 
the sum of (\ref{untwisted}) and (\ref{twisted}) are found to be of the form
$I_{GFF} = 1/2(2 \pi)^3 \sum_{i,a} b^i_a \, G_i \wedge {F_a}^n \wedge {F_a}_n$ 
and $I_{GRR} = 1/48(2 \pi)^3 \sum_{i} b^i 
\, G_i \wedge R^{\mu \nu} \wedge R_{\mu \nu}$, with coefficients 
$b^i_a$ and $b_i$ given by:
\bea
&& b^i_a = - C(G_a) + \sum_{R_a} T(R_a) (1 + 2 n^i_{R_a}) \;,\\
&& b^i = 21 + 1 + \delta_T - {\rm dim}(G) 
+ \sum_{\alpha} (1 + 2 n^i_{\alpha}) \;.
\eea
The correct modular weight $n_i$ is automatically obtained for each 
state. In $b^i_a$, the first contribution is from the gauge bosons and 
involves the quadratic Casimir $C(G_a)$ of the relevant group factor $G_a$, 
whereas the sum in the second term runs over charged states with non-trivial 
representations $R_a$ with respect to the gauge group factor $G_a$,
with character $T(G_a)$. In $b^i$ instead, the $21$ corresponds to the 
gravitino, the $1$ to the dilatino, the quantity $\delta_T$ to the untwisted 
moduli, and the sum in the last term is over all matter states $\alpha$.
The above expressions for the coefficients $b^i_a$ and $b_i$ coincide term 
by term with the well-known results of \cite{dfkz,il}, for each single 
contribution, showing perfect agreement for mixed anomalies

In the case of pure target-space anomalies, instead, the string-derived 
results extracted from (\ref{untwisted})-(\ref{twisted}) and those expected 
from the supergravity analysis of \cite{bg2} differ quite radically in the 
untwisted sector, although agreement is found for twisted sectors. 
Not only the single contributions from each untwisted state disagree, but 
they also sum up to different total results, thus showing a true discrepancy. 

The anomaly we find therefore differs concretely from the one expected
from previous analyses in the literature, beyond the terms linear in 
target-space curvatures. The difference could be related to subtleties 
associated to the compositeness of the target-space connection, 
which could invalidate the analogy with a gauge connection when many of them 
occur as external states. As a matter of fact, it turns out that the total 
anomaly derived from our string computation nicely factorizes, whereas
we have checked that the total anomaly computed by assuming that the 
target-space dependence is analogous to the gauge dependence does not 
factorize. We take this as an indirect argument in favour of our results,
and therefore assume that they should be reproduced by a more careful 
supergravity analysis.

The two explicit examples that we have analysed in detail exhibit a 
certain number of properties that we expect to be quite general,
and valid for any orbifold model free of $U(1)$ anomalies and fixed planes.
Defining the quantity ${\rm tr}F^2$ to represent the appropriate sum of 
terms over the various group factors $G_a$ arising in a given model: 
${\rm tr}F^2 = 1/2 \sum_a {F_a}^n \wedge {F_a}_n$, the anomaly can be 
written in a universal way as 
$I_{FT} = X_2 \wedge X_4$ with:
\bea
&& X_2 = \frac {15}{2(2 \pi)^3} \Big(\sum_{i=1}^3 G_i \Big) \;, 
\label{X2} \\
&& X_4 = {\rm tr} \, R^2 - {\rm tr} \,F^2 + 2 \sum_{i=1}^3 G_i^2 \;.
\label{X4}
\eea
These same quantities enter also the GS term $-2\pi B \wedge X_2$ and the 
modified Bianchi identity $d H = X_4$.
These general results encode full information about all possible
target-space, gauge and gravitational anomalies in four-dimensional 
heterotic vacua without anomalous $U(1)$'s and threshold corrections,
as well as the GS mechanism through which they are cancelled.
The pure target-space part is a novel result, at both the field-theory 
and the string-theory level. The generalization to models with an anomalous
$U(1)$ does not present any technical problem, and only the two-form 
$X_2$ entering the GS term is expected to be modified, through 
additional terms proportional to the field strength of the anomalous
$U(1)$. Models with fixed planes and threshold corrections, instead, 
do present interesting new features \cite{dkl}.

The details of the GS mechanism of anomaly cancellation occurring at the level 
of low-energy effective action do not present any qualitative novelty. 
As usual, the cancellation mechanism can be understood both in a linear 
and a chiral multiplet description \cite{bg1,dfkz}. In the linear basis, the 
part of the effective Lagrangian relevant to anomaly cancellation is
given by
\be
L_l = - \frac 1{12} |d B - X_4^{(0)}|^2 - 2 \pi B \wedge X_2 \;.
\label{Ll}
\ee
The modified kinetic term of $B$ requires that $\delta B = X_4^{(1)}$ under 
gauge, gravitational or target-space transformations. The GS term
induces an anomaly $\delta L_l = - 2 \pi X_2 \wedge X_4^{(1)}$, which is 
equivalent to the WZ descent of $I_{GS} = - X_2 \wedge X_4$, modulo 
irrelevant local terms. This exactly cancels the field-theory anomaly 
$I_{FT} = X_2 \wedge X_4$. In the chiral basis, the relevant Lagrangian 
for the axion field $\chi$ is instead:
\be
L_c = \frac 12 |d \chi - X_2^{(0)}|^2 - 2 \pi \chi X_4 \;.
\label{Lc}
\ee
The shift in the kinetic term now involves only the target-space connection.
This requires $\delta \chi = X_2^{(1)}$ under target-space transformations. 
The GS term induces an anomaly $\delta L_c = - 2 \pi X_2^{(1)} \wedge X_4$, 
which is again equivalent to the WZ descent of $I_{GS} = - X_2 \wedge X_4$,
cancelling as before the field-theory anomaly $I_{FT} = X_2 \wedge X_4$.
In both approaches, one can easily generalize the above results for the CP-odd 
part to full superspace expressions, encoding for instance also CP-even
target-space anomalies.

\section{Geometric structure and relation to torsion}

The general structure of the anomalies derived in the previous sections 
can be understood quite nicely from a geometrical point of view, and 
reveals some interesting features. More precisely, it is possible to 
understand both eqs. (\ref{X2}) and (\ref{X4}) from a ten-dimensional 
point of view.

Let us begin by recalling some well-known facts about the ten-dimensional 
case. The tree-level effective Lagrangian of the $D=10$ heterotic theory 
exhibits a modified kinetic term for the NSNS two-form $B$, involving the 
field-strength $H = d B - Y_4^{(0)}$, where the Chern--Simons three-form 
$Y_4^{(0)}$ is defined as the WZ descent of the four-form
\be
Y_4 = {\rm tr}\, R^2 - {\rm tr}\, F^2 \;.
\label{Y4}
\ee
Correspondingly, the NSNS field has a modified Bianchi identity
$d H = Y_4$ and transforms non-trivially as $\delta B = Y_4^{(1)}$
under gauge transformations or diffeomorphisms. The well-known GS mechanism
then occurs through the variation of the string-generated one-loop 
effective coupling $- 2 \pi B \wedge Y_8$, where the eight-form $Y_8$ is 
given by 
\be
Y_8 = \frac 18 {\rm tr}\, R^4 + \frac 1{32} ({\rm tr}\, R^2)^2 
+ \frac 14 ({\rm tr}\, F^2)^2 
- \frac 18 {\rm tr}\, F^2 {\rm tr}\, R^2 \;.
\label{Y8}
\ee
An important thing to note is that the two-form $R$ entering both 
(\ref{Y4}) and (\ref{Y8}) is in fact a generalized curvature hiding 
also a dependence on the torsion $H$\footnote{The Bianchi identity 
for $H$ then must be understood and solved iteratively, 
in an $\alpha^\prime$ expansion.}. Indeed, it has been shown in 
\cite{hull2} that various requirements, such as space-time supersymmetry 
and world-sheet conformal invariance, fix unambiguously the 
connection $\omega$ which defines $R$ in the above equations to be the 
sum of the usual spin connection and the torsion connection constructed
from $H$. This is also clear from a direct computation of the anomaly 
along the lines of section 3, which is easily generalized to include the 
effect of torsion through vertex operators for the $B$ field.

Let us now consider an orbifold compactification leading to $D=4$.
The spectrum of the states arising from the untwisted spectrum can 
easily be deduced starting from $D=10$. In particular, the $D=10$ 
graviton and NSNS two-form $B$ give rise (in addition to the 
$D=4$ graviton, two-form $B$ and possibly other scalars) to the
three diagonal untwisted complex fields (\ref{Tdef}).
In standard orbifold compactifications, the internal metric and NSNS
flux are fixed, and the vacuum expectation value of the $T$-moduli is 
constant, implying in particular that there is no background torsion in 
the internal manifold. However, a non-trivial generalized connection 
is induced by the fluctuations of the complexified internal metric 
represented by the untwisted $T$-moduli. One finds
\be
\omega_{\mu}^{\;\; i\bar \imath } = Z_{\mu i} \;,
\label{om-Z}
\ee
where $Z_i$ are precisely the target-space connections (\ref{Zi}). 
Correspondingly, the internal components of the 
generalized curvature two-form $R$ do not vanish, but rather 
$R_{i \bar \imath} = G_i$, and the ten-dimensional quantity ${\rm tr}R^2$ 
decomposes into its four-dimensional analogue, plus a contribution 
coming from the internal components: 
\be
{\rm tr}_{D=10} R^2 = {\rm tr}_{D=4} R^2 + 2 \sum_{i=1}^3 G_i^2 \;.
\ee
The general structure (\ref{X4}) of $X_4$ in $D=4$ is hence completely
fixed by the dimensional reduction of the well-known expression (\ref{Y4}) 
for $Y_4$ in $D=10$.
A similar interpretation can be given also for $X_2$ in (\ref{X2}). 
Indeed, the two-form $X_2$ entering the GS term in a $D=4$ model can be
obtained by integrating the $D=10$ GS term $Y_8$ given by (\ref{Y8}) over 
the compactification manifold. Much in the same way as $U(1)$ terms in 
$X_2$ comes from couplings in (\ref{Y8}) with one gauge curvature in 
the four-dimensional directions \cite{w,dsw}, target-space terms 
can arise from the torsion dependence in (\ref{Y8}).

Using the geometric interpretation just described, one can also obtain a 
much better understanding of the various characteristic classes 
appearing in the contribution of the various chiral fields to 
(\ref{untwisted}) and (\ref{twisted}).
Again, as far as untwisted states are concerned, it is convenient and 
justified to reinterpret the result from a $D=10$ point of view, in which
target-space transformations (\ref{sl2r}) are directly related to 
internal reparametrizations of the orbifold.
The first line in (\ref{untwisted}) clearly comes from a chiral gravitino 
in $D=10$, which gives rise, when dimensionally reduced to $D=4$, to a 
chiral gravitino transforming as an internal spinor (first term), plus 
chiral spinors transforming as an internal gravitino (second term), 
with respect to the reparametrizations of the orbifold induced by 
(\ref{sl2r}). The second and third lines in (\ref{untwisted}), instead, 
come from charged $D=10$ chiral spinors, the gauginos, which give 
rise to multiplets of chiral spinors in $D=4$ transforming as internal 
spinors. Twisted states arise from given fixed points of the orbifold and 
therefore cannot be understood from a $D=10$ point of view.
It is clear from the form of (\ref{twisted}) that 
they are simply $D=4$ spinors and transform just as $U(1)$ charged 
fields under reparametrizations of the orbifold, with charges given by 
the modular weights $n_i$. As already noticed in section 5, the 
reason for the different structure of anomalies in untwisted and twisted 
sectors is related to the very different way in which the corresponding 
states arise geometrically. 

\section{Conclusions}

We have shown that target-space anomalies in heterotic orbifold models 
can be understood on the same footing as gauge and gravitational anomalies,
allowing for a more general and unified analysis of them in such 
models. 
All anomalies cancel through a GS mechanism mediated by the
dilaton multiplet, thanks to a new term in the Bianchi identity
for the NSNS $B$-field that involves the target-space curvature.
The key property behind this cancellation is again modular invariance, 
and both the quantum anomaly and the classical GS term are encoded in 
a generalized elliptic genus with well-defined modular properties. 
Although we have focused on the simplest four-dimensional orbifold models,
it is clear that the validity of these results extends straightforwardly 
to a much wider class of models, involving for example non-standard embedding,
Wilson lines, or a different number of space-time dimensions. 
A similar analysis should also be possible for models with fixed planes 
and threshold corrections in $N=2$ sectors.

Several interesting novelties have emerged in the low-energy understanding
of target-space symmetry as well. For instance, a precise relation
to torsion has emerged, which leads to a nice geometric interpretation.
Our results for quantum anomalies involving more than one target-space
connection disagree from the naive field theory expectation based on the
analogy with a gauge connection, suggesting interesting technical subtleties
to be unravelled. 

A very interesting application of target-space symmetries has recently 
emerged in the context of a possible duality between certain heterotic 
orbifold and Type IIB orientifold $N=1$ vacua in $D=4$ 
\cite{abpss}--\cite{afiuv}. Whereas these
symmetries have a good reason to persist quantum mechanically 
on the heterotic side (the underlying T-duality), they are 
apparently accidental on the orientifold side, and therefore 
constitute a very stringent test of the proposed duality. 
It has been shown recently in \cite{ss4} that one-loop anomalies 
in the simplest $\Z_3$ and $\Z_7$ orientifold models are cancelled 
through a generalized GS mechanism mediated by several twisted RR 
fields, as proposed in \cite{iru2}.
Together, the results derived in this work for heterotic models and 
those derived in \cite{ss4} for Type IIB orientifold models demonstrate 
that at least the simplest vacua in both theories do indeed have the same 
target-space symmetry and are free of any anomaly. In spite of 
other problems pointed out so far in the literature \cite{abd,lln}, this is
quite a suggestive result in favour of the conjectured duality.
Finally, it is interesting to note that the same peculiar structure of 
the target-space anomalies appears in both heterotic orbifolds and Type 
IIB orientifolds.

\acknowledgments

It is a pleasure to thank W. Lerche and A. Uranga for very interesting 
discussions, and G. L. Cardoso for useful comments. 
C.A.S. also thanks the International School for Advanced Studies 
(ISAS-SISSA) for hospitality during the completion of this work.

\appendix

\section{Characteristic classes}

The relevant characteristic classes appearing in the anomaly from chiral 
spinors and Rarita--Schwinger fields are the Roof-genus and the 
$G$-polynomial \cite{agw}, functions of the gravitational curvature $R$ and 
defined in terms of the skew eigenvalues $\lambda_a$ of $R/2\pi$ as: 
\bea
&& \widehat{A}(R) = \prod_{a=1}^{D/2} \frac{\lambda_a/2}
{\sinh \lambda_a/2} \;, \label{Ahat} \\ 
&& \widehat{G}(R) = \prod_{a=1}^{D/2} \frac{\lambda_a/2}
{\sinh \lambda_a/2} \Big(2 \sum_{b=1}^{D/2} 
\cosh \lambda_b \,-\, 1\Big) \;. \label{Ghat}
\eea
For the target-space dependence, similar characteristic classes turn
out to appear; these are functions of the composite curvature $G=dZ$ and 
defined, in terms of $g_i = G_i/2\pi$, as
\bea
\widehat A_k(G) &=& \prod_{i=1}^3
\frac {\sin (\pi k v_i)}{\sin (g_i/2 + \pi k v_i)} \;, 
\label{Ak} \\
\widehat G_k(G) &=& \prod_{i=1}^3
\frac {\sin (\pi k v_i)}{\sin (g_i/2 + \pi k v_i)} 
\Big(2 \sum_{j=1}^{3} \cos (g_j + 2 \pi k v_j) \,-\, 1\Big) \;.
\label{Gk}
\eea
The gauge dependence is influenced by the orbifold projection, which acts 
on the $SU(3)$ part of the maximal gauge group and projects it to 
a subgroup $H$. The element $g^k$ of the orbifold group $\Z_N$ acts as 
$F/2\pi \rightarrow F/2\pi + 2 \pi i k V$ on the rescaled gauge curvature 
of the visible gauge group, where $V$ is an $SU(3)$ shift with 
skew-eigenvalues $v_i$. This shift acts as $y_i \rightarrow y_i + 2 \pi k v_i$ 
for $i=1,2,3$. It is then natural to define the following twisted Chern 
characters:
\be
{\rm ch}^k(F) = {\rm ch}(F + (2 \pi)^2 k V) \;.
\label{chk}
\ee
Finally, it is convenient to define:
\bea
&& {\rm ch}^k_{q_i}(F) = \exp \Big\{i \sum_{i=1}^3 
q_i (y_i + 2 \pi k v_i) \Big\} \;, \label{chq} \\
&& {\rm ch}^k_{n_i}(G) = \exp \Big\{\!-\!i\sum_{i=1}^3(1 + 2 n_i)
(g_i/2 + \pi k v_i)\Big\} \;. \label{chn}
\eea

\section{Decomposition of characters}

We report in this appendix some useful explicit expressions for the 
Chern characters relevant to orbifold models with standard embedding. 
Recall first that the $SO(16)$ group underlying $E_8$ decomposes 
naturally into $SO(6) \times SO(10)$, and the eight skew eigenvalues 
of the rescaled $SO(16)$ curvature $F/2\pi$ ($y_i$, $i=1,...,8$) split 
into the three skew eigenvalues of the $SO(6)$ curvature 
($y_i$, $i=1,...,3$) and the five skew eigenvalues of the $SO(10)$ 
curvature ($y_i$, $i=4,...,8$). The internal $SO(6) \sim SU(4)$ group 
further decomposes into $U(1) \times SU(3)$, the rescaled $U(1)$ curvature 
being $\sum_i y_i/6$. For standard embeddding, one of the two $E_8$ groups 
of the $D=10$ theory is broken to $E_6\times H$ in $D=4$, where the $E_6$ 
group is constructed from $SO(10) \times U(1)$ and $H$ is a subgroup of 
$SU(3)$. The Chern characters of the relevant $E_8$ and $E_6$ representations 
can then be derived using their decomposition with respect to the defining 
$SO(16)$ and $SO(10)$ subgroups respectively. 

The basic fundamental and positive/negative spinor representations of 
$SO(2n)$ are easily computed. One finds:
\bea
&& {\rm ch}_{\bf 2n}(F) = 2 \sum_{i=1}^n \cos y_i \;, \\
&& {\rm ch}_{\bf 2^{n-1}}(F) = 2^{n-1} 
\Big(\prod_{i=1}^n \cos y_i/2 \pm \prod_{i=1}^n i \sin y^i/2 \Big) \;,
\label{so1}
\eea
in terms of the skew-eigenvalues $y_i$ of the field strength $F/2 \pi$.
The Chern character in the adjoint representation is then obtained as
\be
{\rm ch}_{{\bf n(2n-1)}}(F) = \frac 12 
\left[ {\rm ch}^2_{\bf 2n}(F) - {\rm ch}_{\bf 2n}(2F) \right] \nn \;.
\label{so2}
\ee

The adjoint representation ${\bf 248}$ of $E_8$ decomposes as 
${\bf 248} = {\bf 120} \oplus {\bf 128}$ under $SO(16)$. 
Its Chern character is then found to be
\be
{\rm ch}_{\bf 248}(F) = 8 + 128 \, \Big(\prod_{i=1}^8 \cos y_i/2
+ \prod_{i=1}^8 \sin y_i/2 \Big) 
+ 2 \!\sum_{i \neq j = 1}^8\! \cos y_i \cos y_j \;. \nn \\
\label{248}
\ee
The relevant representations of $E_6$ and $SU(3)$ are those appearing 
in the decomposition of the adjoint of $E_8$ under the maximal subgroup 
$E_6 \times SU(3)$:
\be
{\bf 248} \rightarrow ({\bf 78}, {\bf 1}) \oplus ({\bf 1}, {\bf 8}) \oplus
({\bf 27}, {\bf 3}) \oplus(\overline{\bf 27}, \bar{\bf 3}) \;.
\label{E8rep}
\ee
The adjoint ${\bf 78}$ and ``fundamental'' ${\bf 27}$ representations 
of $E_6$ further decompose as follows with respect to the 
$SO(10)\times U(1)$:
\bea
{\bf 78} &\rightarrow& 
{\bf 45}_0 \oplus  {\bf 16}_3 \oplus  \overline{{\bf 16}}_{-3} 
\oplus {\bf 1}_0 \;, \nn \\ 
{\bf 27} &\rightarrow& 
{\bf 16}_{-1} \oplus  {\bf 10}_2 \oplus {\bf 1}_{-4} \;. \nn
\eea
The $U(1) \times SU(3)$ part of the Chern characters is easily obtained
by returning to the original $SO(6)$ representations.
For example, using the decomposition of the ${\bf 4}$ of $SO(6)$ as 
${\bf 1_3} \oplus {\bf 3_{-1}}$ under $U(1) \times SU(3)$ and applying 
(\ref{so1}), one can obtain the Chern characters in the ${\bf 1_3}$
and the ${\bf 3_{-1}}$ representations. Generalizing to an arbitrary 
$U(1)$ charge, one finds
\bea
{\rm ch}_{\bf 1_q}(F) &=& \exp \Big(iq \sum_{i=1}^3 y_i/6\Big) \;, \nn \\ 
{\rm ch}_{\bf 3_{q}}(F)&=& \exp \Big(i(q-2)\sum_{i=1}^3 y_i/6\Big) 
\sum_{i=1}^3 \exp i y_i \;. \nn
\eea
Similarly, the Chern character of the adjoint representation ${\bf 8}$ of 
$SU(3)$ is easily deduced from the decomposition 
${\bf 3} \otimes {\bf \bar 3} = {\bf 8} \oplus {\bf 1}$.
The explicit form of the Chern characters for the remaining relevant 
representations is then found to be:
\bea
&& {\rm ch}_{({\bf 78}, {\bf 1})}(F) = 6 + 32 \, \Bigg[
\cos \Big(\sum_{i=1}^3 y_i/2\Big) \prod_{j=4}^8 \cos y_j/2
+ \sin \Big(\sum_{i=1}^3 y_i/2\Big) \prod_{j=4}^8 \sin y_j/2 \Bigg] \nn \\
&& \hspace{68pt} +\, 2 \!\sum_{j \neq k = 4}^8\! \cos y_j \cos y_k \;, 
\label{78} \\
&& {\rm ch}_{({\bf 1}, {\bf 8})}(F) = 2 
+ \!\sum_{i \neq j = 1}^3\! \cos (y_i - y_j) \;, \label{8} \\
&& {\rm ch}_{({\bf 27}, {\bf 3})}(F) = \Bigg[\exp 
\Big(-i\sum_{i=1}^3 y_i\Big) + 16 \exp \Big(-i\sum_{i=1}^3 y_i/2\Big) 
\Big(\prod_{j=4}^8 \cos y_j/2 + \prod_{j=4}^8 \sin y_j/2 \Big) \nn \\
&& \hspace{72pt} +\, 2 \sum_{j = 4}^8 \cos y_j \Bigg] 
\sum_{i=1}^3 \exp i y_i \;. \label{273}
\eea

Notice finally that the decomposition (\ref{E8rep}) can be technically 
understood on the Chern characters (\ref{248}), (\ref{78})--(\ref{273}) 
by using the following trigonometric identity
\bea
&& \prod_{i=1}^3 2 \sin y_i/2 = 
\sum_{i=1}^3 2 \sin \Big(y_i - \sum_{j=1}^3 y_j/2\Big) 
- 2 \sin \Big(\sum_{i=1}^3 y_i/2\Big) \;, \nn \\
&& \prod_{i=1}^3 2 \cos y_i/2 = 
\sum_{i=1}^3 2 \cos \Big(y_i - \sum_{j=1}^3 y_j/2\Big) 
+ 2 \cos \Big(\sum_{i=1}^3 y_i/2\Big) \;, \nn 
\eea
implementing the $SO(6) \rightarrow U(1) \times SU(3)$ decomposition.

\section{Limits of the partition functions}

In this appendix, we report some useful details about the computation 
of the $\tau \rightarrow i\,\infty$ limit of the partition functions
$Z_{ST}(R)$, $Z_{C}^{k,l}(G)$ and $Z_\Gamma^{k,l}(F)$ in eqs. 
(\ref{ZTot1})--(\ref{ZTot3}).

In the untwisted sector, the limits are easily obtained, and one finds:
\bea
&& Z_{ST}(R) \rightarrow q^{-\frac 1{12}} 
\Bigg[\widehat A(R) + \Big(\widehat G(R) -\widehat A(R)\Big) \, q
\Bigg] \;, \\
&& Z_{C}^{k,l=0}(G) \rightarrow \epsilon_k N_k^{-\frac 12} \,
q^{-\frac 14} \Bigg[1 + \Big(\widehat A_k(G) + \widehat G_k(G)\Big)\, 
q\Bigg] \;, \\
&& Z_\Gamma^{k,l=0}(F) \rightarrow q^{-\frac 23}
\Bigg[1 + \Big({\rm ch}_{\bf 248}(F^\prime) 
+ {\rm ch}_{({\bf 78},{\bf 1})}(F) + {\rm ch}^k_{({\bf 1},{\bf 8})}(F) \nn \\
&& \hspace{115pt} +\, {\rm ch}^k_{({\bf 27},{\bf 3})}(F)
+ {\rm ch}^k_{({\bf \overline{27}},{\bf \overline{3}})}(F)
\Big)\, q \Bigg] \;.
\eea
Combining these expressions, the result (\ref{untwisted}) is obtained.

In the twisted sector, the situation is more involved. 
Indeed, there are in general matter states with both vanishing and 
non-vanishing oscillator number $N_L$; correspondingly, there will be a 
complicated interference of terms between $Z_C$ and $Z_\Gamma$ in the limit
$\tau \rightarrow i\, \infty$. 
For this reason, it is convenient to recall how these states 
arise in general. The mass-shell condition for a generic massless state 
in the $l$-th twisted sector reads
\be
N_L - 1 + \frac {\eta}2 + \frac {(p + vl)^2}2 = 0 \;,
\label{mass}
\ee
where $\eta = \sum_{i=1}^3 \theta_i^l(1 - \theta_i^l)$
represents the shift in the zero-point energy in each model and 
takes the value $2/3$ for $\Z_3$ and $4/7$ for $\Z_7$.
Such a state is in general created from the Fock vacuum, with certain 
numbers $m_i$ and $\tilde m_i$ of left- and right-moving bosonic creation 
operators in the $i$-th internal direction, determining an oscillator 
number $N_L$; it is associated with lattice vectors $p$ satisfying 
$(p+v)^2 = 2 - 2 N_L - \eta$. Since this is the most general massless 
twisted state, the relevant terms to keep in $Z_{ST}$, $Z_C$ and $Z_\Gamma$ 
are those of the type $Z_{ST} \sim \bar q^{-\frac 1{12}}$, 
$Z_C \sim \bar q^{-\frac 14 + \frac {\eta}2 + N_L}$ and
$Z_\Gamma \sim \bar q^{-\frac 23 + \frac {(p + v)^2}2}$. Indeed, the 
corresponding term in the total partition function is then
$Z \sim q^{N_L - 1 + \frac {\eta}2 + \frac {(p + v)^2}2}$, and
represents a massless contribution whenever the mass-shell condition 
(\ref{mass}) is satisfied. One can therefore restrict from the beginning 
to these types of terms when evaluating the partition functions for 
$\tau \rightarrow i\,\infty$. 

The limit of the internal partition function $Z_C$ is easily obtained 
by using the infinite sum representation of $\theta$-functions. The 
integer $n$ labelling the infinite sum in the definition of the 
$\theta$-function associated to a given complex direction is related
to the level of the bosonic oscillator that creates the corresponding state,
and positive/negative values of $n$ correspond to left-/right- moving 
oscillators. Since all the massless states occurring in the considered 
models involve at most oscillators of lowest level, it is 
enough to keep the first two subleading terms (in $q$) beside the
leading term, in each $\theta$-function. 
One then finds:
\be
Z_{C}^{k,l \neq 0}(G) \rightarrow - i \, \epsilon_l \,
\Bigg[\bar q^{-\frac 14 + \frac {\eta}2} \,{\rm ch}^k_{-(1-\theta_i^l)}(G)
+ \sum_{s} \, \bar q^{-\frac 14 + \frac {\eta}2 + N_L(s)} 
{\rm ch}^k_{n_i(s)}(G) \Bigg] \;,
\ee
with the following expressions for the oscillator number $N_L$ and the 
modular weight $n_i$ for the generic massless twisted state $s$:
\bea
&& N_L(s) = - \frac 12 \sum_{i=1}^3 
\Big[(m_i - \tilde m_i)(1 - 2 \theta_i^l) - (m_i + \tilde m_i)] \;, 
\label{NL}\\
&& n_i(s) = -(1 - \theta_i^l + m_i - \tilde m_i) \;.
\label{mw}
\eea
Notice that the known expression \cite{il} for the modular weight of a 
generic oscillator state is recovered.

For the lattice partition function, one has to understand which kind
of lattice vectors $p$ can occur for the generic state discussed above, 
and which representation is defined by the corresponding $q=p+v$.
For the $N_L = 0$ states, the allowed $p$'s give rise to a 
${\bf 27}$ of $E_6$ and fixed charges in the $SU(3)$ part, for both 
the $\Z_3$ and $\Z_7$ models. For the states with $N_L \neq 0$, the 
allowed $p$'s lead to three distinct $q$'s in the $SU(3)$ part, which 
we shall label with $q^a$, $a=1,2,3$, giving rise to a triplet for 
$\Z_3$ and three groups of eight singlets for $\Z_7$. More precisely, 
the relevant part of the lattice partition function is found to be:
\be
Z_\Gamma^{k,l \neq 0}(F) \rightarrow 
\bar q^{\frac 13 - \frac {\eta}2} \,
{\rm ch}_{({\bf 27},{\bf 1})} (F) \, 
{\rm ch}^k_{\theta_i^l + \frac {\epsilon_l-3}6} (F)
+ \sum_{a=1}^3 \bar q^{-\frac 23 + (q^a)^2} {\rm ch}^k_{(q^a)_i}(F) \;.
\hspace{0pt}
\ee
One can check that the allowed $q(s)$'s for an oscillator state $s$, 
among the three $q^a$, $a=1,2,3$, are defined by
\be
q^a_i(s) = \theta_i^l - \epsilon_l \delta^a_i + \frac {\epsilon_l - 1}2
\;,\;\; a \in \Big\{a|\epsilon_l \theta_a(l) 
+ \frac {1 - \epsilon_l}2 = N_L(s)\Big\}
\;.
\label{charges}
\ee
Putting everything together, one finally finds
\bea
&& I_{FT}^{\rm tw.}(F,R,G) = 
- \frac i{2N} \sum_{k=0}^{N-1} \sum_{l=1}^{N-1} N_{k,l} \,
\epsilon_l \, \Bigg[{\rm ch}_{({\bf 27},{\bf 1})}(F) \, 
{\rm ch}^k_{\theta_i^l + \frac {\epsilon_l-3}6} (F) \,
{\rm ch}^k_{-(1-\theta_i^l)}(G) \hspace{35pt} \\
&& \hspace{190pt} + \sum_{s} {\rm ch}^k_{q_i(s)}(F)\,
{\rm ch}^k_{n_i(s)}(G) \Bigg] \, \widehat A(R) \;, \nn
\eea
where the sum over the set of oscillator states $s$ now contains also the 
sum over the allowed $q$'s. The charges $q_i(s)$ and modular weights $n_i(s)$
are given by (\ref{mw}) and (\ref{charges}) respectively. Observe finally
that to each particle $s$ in the $l$-th twisted sector there corresponds an 
antiparticle $s^\prime$ in the $(N-l)$-th conjugate sector with left- 
and right-moving oscillators exchanged and opposite charge vector. 
Since $N_L(s^\prime) = N_L(s)$, $n_i(s^\prime) = 1 - n_i(s)$ and 
$q_i(s^\prime) = -q_i(s)$, these two states $s$ and $s^\prime$ give 
identical contributions to the anomaly, as expected. 
Pairing these contributions, the sums over $l$ can be reduced to $l \in S_l$.
The sum over $k$ can be done explicitly by noticing that the $k$-dependence 
hidden in the twisted Chern characters amounts to the phase 
$\exp 2 \pi k i \sum_i v_i (q_i - n_i)$, which can be checked case by case 
to be simply $1$. One then obtains eq. (\ref{twisted}).

\section{Chiral determinants}

In this appendix, we provide some details about the evaluation of the 
determinants appearing in section 3. In particular, we show how two 
different regularizations -- modular-invariant and holomorphic -- give 
different results: the elliptic genera $\bar A(\tau;F,R,G)$ and $A(\tau;F,R,G)$
respectively. The typical quantity to be computed is a chiral determinant
with periodicities $\alpha,\beta \in [0,1[$ and a twist 
$\lambda$\footnote{In our case, the twist $\lambda$ is always a given 
two-form, so the expressions that will follow are formal and eventually 
have to be understood as an expansion in $\lambda$.}:
\be
{\rm det}_{\alpha,\beta} (\lambda ) = \prod_{m,n} \frac{2\pi}{\tau_2} \,
\Big[(m-\beta) - (n+\alpha) \, \tau + \lambda \Big] \;.
\label{det}
\ee
It is well known that chiral determinants such as (\ref{det}) are 
ambiguous. A possible approach is to multiply (\ref{det}) by its complex 
conjugate, which corresponds to the opposite chirality. However,
although the square modulus of (\ref{det}) can be computed, it fails 
to factorize into the product of a holomorphic and an anti-holomorphic parts 
associated to components with opposite chirality.
In order to define in an unambiguous way the original chiral 
determinant, one can start with $\lambda$ and $\bar \lambda$ independent
of each other, {i.e.} different twists in the chiral and antichiral parts,
and define the original chiral determinant as 
\be
{\rm det}_{\alpha,\beta} (\lambda ) \equiv 
\frac {\Big|{\rm det}_{\alpha, \beta}(\lambda )\Big|^2} 
{{\rm det}_{\alpha, \beta}(0)}\Bigg|_{\bar \lambda=0}\;.
\label{det-def}
\ee
It is worth observing that this definition precisely corresponds to the 
original form of the determinants yielding the topological generating 
functional $\bar A$. In each space-time direction there is a bosonic 
determinant with left twist $\lambda \neq 0$ and right twist 
$\bar \lambda = 0$, and a fermionic determinant with $\lambda = 0$.

In the non-degenerate case $(\alpha,\beta)\neq (0,0)$, the square modulus 
of (\ref{det}) can be computed from the representation \cite{pol}:
\bea
&& \ln \Big|{\rm det}_{\alpha,\beta} (\lambda)\Big|^2 = 
\frac 12 \lim_{s\rightarrow 0} \frac d{ds} \sum_n \Bigg[\int_{C_+}\! dz 
\Big(\frac{e^{i\pi z}}{i\sin \pi z} f_{\alpha,\beta}^{-s}(z,\lambda) 
- f_{\alpha,\beta}^{-s}(z,\lambda) \Big) \nn \\
&& \hspace{153pt} + \int_{C_-}\! dz \Big( \frac{e^{-i\pi z}}{i\sin \pi z}
f_{\alpha,\beta}^{-s}(z,\lambda) + f_{\alpha,\beta}^{-s}(z,\lambda)\Big) 
\Bigg] \;, \hspace{30pt} \label{lndet}
\eea
where
\be
f_{\alpha,\beta}(z,\lambda)= (\frac{2\pi}{\tau_2})^2 
\Big[(z-\beta) - (n+\alpha) \, \tau + \lambda \Big]
\Big[(z-\beta) - (n+\alpha) \, \bar\tau + \bar \lambda \Big] \;.
\ee
We have performed a Sommerfeld--Watson transformation to convert the sum 
over $m$ to an integral over closed contours $C_\pm$ following the real 
axis and going at infinity in the upper/lower half-plane.
The first and third terms in the square brackets converge for 
$s\rightarrow 0$, so that the derivative at $s=0$ can be taken directly.
This gives rise to $\ln f_{\alpha,\beta}(z,\lambda)$, which has branch 
points in the complex plane at the locations  
$z_+ = \beta -\lambda + (n+\alpha) \, \tau $ 
and $z_- = \beta - \bar \lambda + (n+\alpha) \, \bar \tau $.
The corresponding integrals can then easily be performed, taking care
of deforming the $C_{\pm}$ contours to avoid the branch points.
Their contribution to $\ln |{\rm det}_{\alpha,\beta} (\lambda)|^2$ is
\bea
\sum_{n=0}^{\infty} \ln \Big| 1-q^{n+\alpha} e^{2\pi i(\beta -\lambda)} 
\Big|^2 + \sum_{n=1}^{\infty} \ln \Big|1-q^{n-\alpha} 
e^{-2\pi i (\beta -\lambda)} \Big|^2 \;.
\label{osci}
\eea
Let us now consider the second and fourth terms in the square brackets
in (\ref{lndet}). These terms are convergent only for ${\rm Re}\, s>1/2$,
so we compute the integrals in this regime and then continue analytically
the result to $s=0$. In this case
\bea
&& \int_{C_+}\! dz \, f_{\alpha,\beta}^{-s}(z,\lambda) - 
\int_{C_-}\! dz \, f_{\alpha,\beta}^{-s}(z,\lambda) = 
2 \int_{-\infty}^{+\infty} \! dx \, f_{\alpha,\beta}^{-s}(x,\lambda) \nn \\
&& \hspace{120pt}=  2\Big[(n+\alpha) \tau_2 - {\rm Im}\,\lambda \Big]^{1-2s} 
\frac{\Gamma(1/2)\Gamma(s-1/2)}{\Gamma(s)} \;.
\label{intx}
\eea
The sum over $n$ can then be performed using the $\zeta$-function 
regularization. In this way, one finds a contribution to 
$\ln |{\rm det}_{\alpha,\beta} (\lambda)|^2$ equal to
\be
-2\pi\, \tau_2 \, B_2(\alpha - {\rm Im}\, \lambda/\tau_2) \;,
\label{intxf}
\ee
where $B_2(c)=1/6 +c^2 - c$ is the second Bernoulli 
polynomial\footnote{Strictly speaking, the result (\ref{intxf})
is valid only for twists $\alpha<1/2$ (for sufficiently small $\lambda$).
However, by considering the equivalent $(1-\alpha)$-twist the result 
(\ref{intxf}) can actually be extended to all twists $0<\alpha <1$.}.
By combining (\ref{intxf}) and (\ref{osci}), one finally obtains
\bea
\Big|{\rm det}_{\alpha,\beta} (\lambda)\Big|^2 &=& 
e^{-\frac{2\pi ({\rm Im}\,\lambda)^2}{\tau_2}}\, 
e^{-2 \pi {\rm Im}\,\lambda (1 - 2 \alpha)} \, 
|q|^{\frac 16 + \alpha^2 - \alpha} \nn \\ &\;& \times 
\prod_{n=0}^\infty \Big|1-q^{n+\alpha} e^{2 \pi i (\beta - \lambda)}\Big|^2
\prod_{n=1}^\infty \Big|1-q^{n-\alpha} 
e^{- 2 \pi i (\beta - \lambda)}\Big|^2 \;. \nn
\eea
As anticipated, this expression fails to factorize into holomorphic and 
and anti-holomorphic parts, but the definition (\ref{det-def}) yields
\be
{\rm det}_{\alpha,\beta}(\lambda) =
\frac{\theta_1{\alpha \brack \beta}(\lambda|\tau)}{\eta(\tau)}\,
e^{\frac{\pi \lambda^2}{2\tau_2}} \;.
\label{D1}
\ee

In the degenerate case $(\alpha, \beta)=(0,0)$, the analysis is similar, 
but one has to introduce an IR regulator \cite{pol}. Following the same 
steps as before, and excluding the zero-mode, one then finds
$$
\Big|{\rm det}^{\prime}_{0,0} (\lambda)\Big|^2 = (2 \pi \lambda)^{-2}
e^{-\frac{2\pi {\rm Im}\,\lambda^2}{\tau_2}}\, 
e^{-2 \pi {\rm Im}\,\lambda}\, |q|^{\frac 16}\! 
\prod_{n=0}^\infty \Big|1-q^{n} e^{-2 \pi i \lambda} \Big|^2
\prod_{n=1}^\infty \Big|1-q^{n} e^{2 \pi i \lambda} \Big|^2 \;,
$$
and hence
\be
{\rm det}_{0,0}^{\prime}(\lambda)=
\frac{\theta_1(\lambda|\tau)}{2\pi \lambda\,\eta(\tau)}
e^{\frac{\pi \lambda^2}{2\tau_2}} \;.
\label{D2}
\ee
Note that this result for the degenerate case can be verified by an 
alternative method, which consists in rewriting the determinant in terms 
of Eisenstein series, for which there is a well-known modular-invariant 
regularization (see for example \cite{lnsw}).

The above results for the determinants give rise to the modular-invariant 
generalized elliptic genus $\bar A (\tau;F,R,G)$, eq. (\ref{modinv}).  
On the other hand, a holomorphic regularization 
can be easily obtained, for example, by defining the effect of the 
twist $\lambda$ by analytic contribution from the twists $\alpha$ and 
$\beta$. In other words, the twisted determinant is defined as the 
value of the ordinary determinant for twist 
$(\alpha^\prime,\beta^\prime) = (\alpha,\beta + \lambda)$. 
Proceeding in this way, one simply finds the same results as 
before, (\ref{D1}) and (\ref{D2}), but without the non-holomorphic 
exponential term. Without this term, modular invariance is lost, but the 
results are manifestly holomorphic in $\tau$. The resulting  
elliptic genus is then $A(\tau;F,R,G)$.

The impossibility of having at the same time holomorphicity and
modular invariance is called Quillen's anomaly \cite{quillen}; 
it plays a crucial role in anomaly cancellation in heterotic theories.

\end{document}